%% file: ANA-STDM-2018-39-PAPER.tex
\pdfoutput=1
\documentclass[cernpreprint,texlive=2023,atlasdraft=false,UKenglish, texmf, orcidlogo]{atlasdoc}
 
\usepackage{atlaspackage}
\usepackage{atlasphysics}
\usepackage{atlasbiblatex}
 
\newcommand{\AtlasCoordFootnote}{ATLAS uses a right-handed coordinate system
with its origin at the nominal interaction point (IP) in the centre of the
detector and the $z$-axis along the beam pipe. The $x$-axis points from
the IP to the centre of the LHC ring, and the $y$-axis points upwards.
Cylindrical coordinates $(r,\phi)$ are used in the transverse plane, $\phi$
being the azimuthal angle around the $z$-axis. The transverse energy is
defined as $E_{\mathrm{T}}=E\sin\theta$, where $E$ is the energy and
$\theta$ is the polar angle. The pseudorapidity is defined as
$\eta=-\ln\tan(\theta/2)$ and the angular distance is measured in units
of $\Delta R\equiv\sqrt{(\Delta\eta)^{2}+(\Delta\phi)^{2}}$.
}
 
\graphicspath{{logos/}}
 
\usepackage{ANA-STDM-2018-39-PAPER-defs}

\AtlasTitle{Inclusive-photon production and its dependence on photon
isolation in $pp$ collisions at $\sqrt s=$ 13 TeV using 139 fb$^{-1}$ of ATLAS data}
 
\author{The ATLAS Collaboration}
 
\AtlasJournalRef{JHEP 07 (2023) 086}
\AtlasDOI{DOI:10.1007/JHEP07(2023)086}
\PreprintIdNumber{CERN-EP-2022-247}
 
\AtlasAbstract{
Measurements of differential cross sections are presented for
inclusive isolated-photon production in $pp$ collisions at a
centre-of-mass energy of 13 TeV provided by the LHC and using
139 fb$^{-1}$ of data recorded by the ATLAS experiment. The cross
sections are measured as functions of the photon transverse energy in
different regions of photon pseudorapidity. The photons are required
to be isolated by means of a fixed-cone method with two different cone
radii. The dependence of the inclusive-photon production on the photon
isolation is investigated by measuring the fiducial cross sections as
functions of the isolation-cone radius and the ratios of the
differential cross sections with different radii in different regions
of photon pseudorapidity. The results presented in this paper
constitute an improvement with respect to those published by ATLAS
earlier: the measurements are provided for different isolation radii
and with a more granular segmentation in photon pseudorapidity that
can be exploited in improving the determination of the proton parton
distribution functions. These improvements provide a more in-depth
test of the theoretical predictions. Next-to-leading-order QCD
predictions from JETPHOX and SHERPA and next-to-next-to-leading-order
QCD predictions from NNLOJET are compared to the measurements, using
several parameterisations of the proton parton distribution
functions. The measured cross sections are well described by the
fixed-order QCD predictions within the experimental and theoretical
uncertainties in most of the  investigated phase-space region.
}

\hypersetup{pdftitle={ATLAS document},pdfauthor={The ATLAS Collaboration}}
 
\begin{document}
 
\maketitle
 
\tableofcontents
 
\newpage
 
\section{Introduction}
The production of prompt photons\footnote{Photons that are not
secondaries from hadron decays are considered as prompt.} at high
transverse momentum ($\pt$) in proton--proton collisions, $\ppg$,
provides a testing ground of perturbative QCD (pQCD) in a cleaner
environment compared to jet production, since it is less affected by
hadronisation effects.  At leading order (LO) in pQCD, two processes
contribute to prompt-photon production: the direct process, in
which the photon originates directly from the hard interaction, and
the fragmentation process, in which the photon is produced when
a high $\pt$ parton fragments~\cite{pr:d76:034003,pr:d79:114024}. In
hadron colliders, photons are produced copiously in decays of neutral
hadrons; thus, isolation requirements are necessary to separate
prompt-photon production,
whose dynamics is governed by pQCD, from those photons arising
from hadron decays. The inclusive production of isolated photons in
$pp$ collisions has been studied previously by
ATLAS~\cite{pr:d83:052005,pl:b706:150,pr:d89:052004,jhep:1608:005,pl:b770:473,jhep:10:203}
and CMS~\cite{prl:106:082001,pr:d84:052011,epj:c79:20} at
centre-of-mass energies (\!\!$\sqrt s$) of $7$, $8$ and $13$~\TeV.
 
This paper presents measurements of inclusive isolated-photon
production in $pp$ collisions at $\sqrt s=13$~\TeV\ with the ATLAS
detector at the LHC using an integrated luminosity of $139$~fb$^{-1}$
collected between 2015 and 2018. Differential cross sections as
functions of the photon transverse
energy,\footnote{\AtlasCoordFootnote} $\etg$, are measured in
different regions of the photon pseudorapidity, $\etag$, for
$\etg>250$~\GeV\ and $|\etag|<2.37$. The photon is required to be
isolated at particle level by demanding that the transverse energy of
the stable particles within a cone of
radius $R=0.4$ or $R=0.2$ around the photon direction, $\etiso$, is
smaller than a certain value; this isolation method is called
`fixed-cone' and
$\etiso<\etisoc\equiv$~\etisocut\ is chosen in this analysis for
the isolation requirement.
 
Next-to-leading-order (NLO) and next-to-next-to-leading-order (NNLO)
pQCD predictions are compared to the measurements. The dominant
production mechanism in $pp$ collisions at
the LHC proceeds via the $qg\rightarrow q\gamma$ process; in this way,
measurements of prompt-photon production are sensitive to the gluon
density in the proton~\cite{np:b860:311,epl:101:61002,pr:742:1} and
can be used as input to global QCD fits to help to constrain the
proton parton distribution functions (PDF). Recent
studies~\cite{epj:c78:470} have shown that the inclusion of
prompt-photon measurements~\cite{jhep:1608:005} from ATLAS provides a
reduction in the gluon density uncertainties.
 
The results presented in this paper extend in several aspects
those at $8$ and $13$~\TeV\ reported in previous
publications~\cite{jhep:1608:005,pl:b770:473,jhep:10:203}. The
measurements use a finer granularity in $\etag$
and so they provide more data points as input
to the QCD fits. The measurements benefit from a reduction of the
experimental systematic uncertainty, especially that in the photon
identification efficiency, as well as from an approximately four-fold
increase in the integrated luminosity.
The dependence of the fiducial cross section on the isolation-cone
radius $R$ is also investigated as well as the ratios of the differential
cross sections for $R=0.2$ and $R=0.4$ as functions of $\etg$ and
$\etag$. These measurements test the $R$ dependence of the inclusive
isolated-photon cross section. At LO pQCD, there is no dependence of
the cross section on $R$ and so the first non-trivial theoretical
contribution arises at higher orders in pQCD~\cite{jhep:04:166}. Therefore, these
measurements provide a test of pQCD at high orders. From the
theoretical point of view, isolation helps to suppress the
fragmentation contribution. The fragmentation component is
available in the calculations from
{\jetp}~1.3.1\_2~\cite{jhep:0205:028,pr:d73:094007} and
\nnlojet~\cite{jhep:08:094}. In the calculations from
{\sher}~2.2.2~\cite{spp:7:034},
an isolation requirement is essential to avoid
divergencies in the matrix elements when the photon is collinear with
a parton. This is achieved by using the method based on the Frixione
criterion~\cite{pl:b429:369}
or
the hybrid method~\cite{jhep:04:166}, which combines the Frixione
criterion and the fixed-cone method. The measurements presented in this paper
are performed using the fixed-cone criterion since, due to the finite
size of the detector elements, a discrete version of the Frixione criterion
leads to large experimental uncertainties.
The $R$ dependence of the measured cross sections allows a test of the
different theoretical approaches to the photon-isolation modelling.
 
The paper is organised as follows: the ATLAS detector is described in
Section~\ref{detector}. The details of the data samples and the Monte
Carlo simulations as well as the event and photon selection are
included in Sections~\ref{datmc} and \ref{evesel}, respectively. The
background evaluation and signal extraction are explained in
Section~\ref{bacsub}: the main background to isolated-photon events
arises from jets misidentified as photons, which includes non-prompt
photons, and is subtracted using a data-driven technique. The strategy
for the cross section measurements is summarised in
Section~\ref{fiducial}. Section~\ref{syst} is devoted to the
description of the experimental uncertainties. Theoretical predictions
and their uncertainties are discussed in Section~\ref{nlo}.  The
results are reported in Section~\ref{res}. A summary is given in
Section~\ref{conc}.

\section{ATLAS detector}
\label{detector}
The ATLAS detector~\cite{PERF-2007-01,ATLAS-TDR-19,Abbott:2018ikt} is
a multipurpose  detector with a forward--backward symmetric
cylindrical geometry. It consists of an inner tracking detector
surrounded by a thin superconducting solenoid, electromagnetic and
hadronic calorimeters, and a muon spectrometer incorporating three
large superconducting toroid magnets. The inner-detector system is
immersed in a $2$~T axial magnetic field and provides charged-particle
tracking in the range $|\eta|<2.5$. The high-granularity silicon pixel
detector is closest to the interaction region and provides four
measurements per track. The pixel detector is followed by the silicon
microstrip tracker, which typically provides four three-dimensional
space point measurements per track. These silicon detectors are
complemented by the transition radiation tracker, which enables
radially extended track reconstruction up to $|\eta|=2$. The
calorimeter system covers the range $|\eta|<4.9$. Within the region
$|\eta|<3.2$, electromagnetic (EM) calorimetry is provided by barrel
and endcap high-granularity lead/liquid-argon (LAr) calorimeters, with
an additional thin LAr presampler covering $|\eta|<1.8$ to correct for
energy loss in material upstream of the calorimeters; for
$|\eta|<2.5$, the EM calorimeter is divided into three layers in
depth. Hadronic calorimetry is provided by a steel/scintillator-tile
calorimeter, segmented into three barrel structures within
$|\eta|<1.7$, and two copper/LAr hadronic endcap calorimeters, which
cover the region $1.5<|\eta|<3.2$. The solid-angle coverage is
completed out to $|\eta|=4.9$ with forward copper/LAr and tungsten/LAr
calorimeter modules, which are optimised for EM and hadronic
measurements, respectively. Events are selected using a first-level
trigger implemented in custom electronics, which reduces the
maximum bunch crossing rate of $40$~MHz to a design value of $100$~kHz
using a subset of detector information. Software algorithms with
access to the full detector information are then used in the
high-level trigger to yield a recorded event rate of about
$1$~kHz~\cite{epj:c77:317}.
 
An extensive software suite~\cite{ATL-SOFT-PUB-2021-001} is used in
the reconstruction and analysis of real and simulated data, in
detector operations, and in the trigger and data acquisition systems
of the experiment.

\section{Data sample and Monte Carlo simulations}
\label{datmc}
\textbf{Data sample.} The data used in this analysis were collected
with the ATLAS detector during the proton--proton collision running
periods from 2015 to 2018, when the LHC operated at a centre-of-mass energy
of $\sqrt s=13$~\TeV. The integrated luminosity of this data set, in
which events are required to pass data quality requirements~\cite{jinst:15:p04003},
is $139.0\pm 2.4$~fb$^{-1}$~\cite{ATLAS-CONF-2019-021}.  Events in
which the calorimeters or the inner detector were not fully
operational or showed data quality problems are excluded.
 
\textbf{Simulated event samples.} Samples of simulated events were produced
using Monte Carlo (MC) techniques to study the characteristics of the
signal events. The MC samples are also used to determine the
ingredients necessary to obtain the measured cross sections. In
addition, MC samples are used to estimate non-perturbative corrections
to the fixed-order pQCD calculations.
 
The MC programs {\pyt}~8.186~\cite{cpc:178:852} and
{\sher}~2.1.1~\cite{jhep:0902:007} were used to generate the simulated
signal events. In both generators, the partonic processes are simulated using
LO matrix elements, with the inclusion of initial- and
final-state parton showers. Fragmentation into hadrons is performed
using the Lund string model~\cite{prep:97:31} in the case of {\pyt},
and a modified version of the cluster model~\cite{epj:c36:381} in the
case of {\sher}. For the samples generated with \pyt\ (\sher), the
proton structure is parameterised using the LO
NNPDF2.3~\cite{np:b867:244} (NLO CT10~\cite{pr:d82:074024}) PDFs.
Both samples include a simulation of the
underlying event (UE). The event generator parameters are set
according to the ``A14''~\cite{ATL-PHYS-PUB-2014-021} tune for {\pyt} and the
tune developed by the authors for use in conjunction with the NLO CT10 PDF
set for {\sher}.
 
The {\pyt} simulation of the signal includes LO matrix elements for
photon plus jet production
from both direct processes (the subprocesses $qg\rightarrow q\gamma$
and $\qq\rightarrow g\gamma$) and photon bremsstrahlung in QCD dijet
events to simulate the fragmentation process. The contribution from
the $qg\rightarrow q\gamma$ subprocess is dominant over most of the
measured phase-space region.
The {\sher} samples are
generated with LO matrix elements for photon plus jet final states
with up to three additional partons.
The photon bremsstrahlung component is simulated differently in \pyt\
and \sher. In \pyt, photons can be radiated in the parton shower
without a restriction on the opening angle with respect to the parent
parton and, as a result, the photons can be emitted very close to the
parton direction. In \sher, photons are not emitted in the parton
shower and the photon bremsstrahlung component is simulated through
matrix elements of $2\rightarrow N$ processes, with $N\geq 3$. In this
case, divergencies in the calculation are avoided by restricting the emission
through an implementation of the Frixione requirement;
as a result, photons are not emitted close to the parent parton.
Frixione's criterion requires the total transverse energy inside a
cone of size ${\mathfrak r}$ in the $\etaphi$ plane around the
generated final-state photon, excluding the photon itself, to be below
a certain threshold, $E_{\mathrm{T}}^{\mathrm{max}}({\mathfrak r})=\epsilon\etg ((1-\cos{{{\mathfrak r}}})/(1-\cos{{\cal R}}))^{n}$, for all ${\mathfrak r} < {\cal R}$, where ${\cal R}$
is the maximal cone size, $n$ is a parameter which modifies the
dependence of the threshold from the radius ${\mathfrak r}$ and
$\epsilon$ is a constant such that $\epsilon\etg$ represents the
threshold for ${\mathfrak r}={\cal R}$. The parameters used for the generation of
these \sher\ samples are chosen to be ${\cal R}=0.3$, $n=2$ and
$\epsilon=0.025$. The resulting Frixione isolation requirement applied
at the generation level in \sher\ is looser than the ones applied in
this analysis at particle and reconstruction levels (see Section~\ref{fiducial}).
 
The second main background after misidentification of jets as photons
arises from electrons or positrons misidentified as photons
and is evaluated using MC samples generated with the program
\sher~2.2.1~\cite{spp:7:034,jhep:12:039,jhep:0202:044,prl:108:111601,jhep:0803:038,jhep:1304:027}.
The $pp \rightarrow Z/\gamma^*\rightarrow e^+e^- + {\mathrm{X}}$
and $pp \rightarrow  W\rightarrow e\nu + {\mathrm{X}}$
processes are generated with matrix elements calculated with up to
two additional partons at NLO and up to four partons at LO. The NNLO
NNPDF3.0 PDF set~\cite{jhep:1504:040} is used in conjunction with a
dedicated set of parton-shower-generator parameters~\cite{spp:7:034}
developed by the \sher\ authors.
 
For all these MC samples, pile-up from additional $pp$ collisions in
the same and neighbouring bunch crossings is simulated by overlaying
each MC event with a variable number of simulated inelastic $pp$
collisions generated using \pyt~8.186 with the ATLAS set of tuned
parameters for minimum bias events (A3
tune)~\cite{ATL-PHYS-PUB-2012-003}. The MC events are weighted
(``pile-up reweighting'') so that the distribution of the average
number of interactions per bunch crossing matches the one observed in data.
All the samples of generated events were passed through the
{\textsc{Geant}}~4-based~\cite{nim:a506:250} ATLAS detector- and
trigger-simulation programs~\cite{epj:c70:823}. The simulated event samples
were reconstructed and analysed by the same program chain as the data.
 
In addition, dedicated MC samples without UE were generated at
particle and parton levels to correct the fixed-order pQCD calculations
for hadronisation and UE effects (see Section~\ref{hadcor}).
 
\section{Event and photon selection}
\label{evesel}
\textbf{Event selection.}
The data sample used consists of events recorded by a single-photon
high-level trigger with a nominal transverse energy threshold of
$140$~\GeV\ and ``loose'' photon identification
requirements~\cite{epj:c77:317,epj:c79:205,epj:c80:47}. The efficiency of the
trigger for photons with $\etg>250$~\GeV\ is found to be close to
$100\%$. The inefficiency of the high-level trigger with respect to
the first-level trigger is found to be subpercent and taken as a
systematic uncertainty (see Section~\ref{rc}).
 
The initial data sample of isolated-photon events is selected offline
from those events recorded by the trigger mentioned above and
by requiring the events to have at least one reconstructed primary
vertex, which has at least two associated tracks of $\pt>500$~\MeV\
and is consistent with the average beam-spot position.
 
\textbf{Photon reconstruction.}
The offline electron- and photon-candidate reconstruction is based on dynamic
variable-size clusters of EM calorimeter cells, called
superclusters~\cite{JINST14:P12006}, which change in size as needed to
recover energy from bremsstrahlung photons or from electrons from
photon conversions. The calibration techniques exploit this advantage
of the dynamic clustering algorithm, while achieving similar linearity
and stability as for the fixed-size clusters used
previously~\cite{epj:c79:205}. Superclusters are based on
topoclusters, which are built using a dynamical topological
cell-clustering algorithm in the three-dimensional
space~\cite{epj:c77:490} and calibrated at the EM scale. An electron
candidate is defined as an object consisting of a supercluster built
from energy deposits in the calorimeter and a matched track. A
converted photon candidate is a supercluster matched to a conversion
vertex (or vertices) or a track consistent with a photon conversion,
and an unconverted photon candidate is a supercluster matched to
neither an electron track nor a conversion vertex. About $20\%$ of
photons at low $|\eta|$ ($|\eta|\lesssim 0.8$) convert in the inner
detector, while up to about $65\%$ of photons convert at  $|\eta|
\approx 2.3$~\cite{JINST14:P12006} due to the non-uniform amount of
material versus $|\eta|$ upstream of the calorimeter.
 
\textbf{Photon calibration}.
The energy calibration of electrons and photons
is updated for the new energy reconstruction~\cite{JINST14:P12006}. The
energy response and resolution of the electrons and photons are
optimised using a multivariate regression algorithm, which exploits
the properties of the cluster energy deposit in the EM
calorimeter. The energy scale corrections extracted from $Z\rightarrow
ee$ decays are applied to correct the photon energy
scale~\cite{JINST14:P12006}. A data-driven validation of the photon
energy scale corrections is performed using radiative decays of the
$Z$ boson, probing the region $\etg\lesssim 80$~\GeV.
The possible nonlinear energy response for higher $\etg$ is covered
by the systematic uncertainties, determined from auxiliary
measurements~\cite{JINST14:P03017}, that are propagated up to
$3$~\TeV\ using MC simulations.
 
Several systematic uncertainties impact the measurement of the energy
of electrons and photons in a way that depends on their transverse
energy, pseudorapidity and, for photons, whether they are
reconstructed as converted or unconverted
candidates~\cite{JINST14:P12006}.
Some of these uncertainties were re-evaluated with respect to the
ones~\cite{JINST14:P03017} used in the previous
publication~\cite{jhep:10:203} to reflect the changes in the
reconstruction described above.
The sensitivity of the calibrated energy to the detector material was
also re-evaluated. The systematic uncertainties due to the material
description of the innermost pixel detector layer and the services of
the pixel detector were also updated using a more accurate description
of these systems.
 
\textbf{Photon identification}.
Photon candidates are identified by using variables that characterise
the lateral and longitudinal electromagnetic shower development in the
EM calorimeter and the energy fraction leaking into the hadronic
calorimeter.
 
The photon identification used in this analysis starts with a
loose selection~\cite{JINST14:P12006}. The signal selection is based on the
``tight''~\cite{JINST14:P12006} photon identification
criteria; tight requirements are imposed on the shower shapes in the
second layer and in the finely segmented first layer of the EM
calorimeter as well as on the energy deposited in the hadronic
calorimeter.
These requirements are optimised separately to ensure the
compatibility of the measured shower profile with that originating
from unconverted or converted photon candidates.
Small differences in the average values of the
shower-shape variables between data and simulation are observed and
corrected for in simulated events  prior to the application of the
photon identification criteria. Non-tight photon candidates, used for
the data-driven background subtraction (see Section~\ref{bacsub}), are
defined as those photons which satisfy the loose criteria, but fail a
given subset of tight requirements~\cite{jhep:10:203}.
 
\textbf{Photon isolation}.
Photon candidates are required to be isolated by using the isolation
transverse energy, $\etiso$. The $\etiso$ variable is constructed by
summing up the transverse energies of all topoclusters within a cone
of radius $R=0.4$ or $R=0.2$ in the $\etaphi$ plane around the photon
cluster barycenter. Only positive energy topoclusters are
used.\footnote{Negative topocluster energies can arise due to the
presence of negative cell signals in the ATLAS calorimeters, which
are the result of fluctuations introduced predominantly by pile-up
and, to a lesser extent, by electronic noise.} The topoclusters
include cells from the EM and hadronic calorimeters. The energy from
the core of the cone in the electromagnetic calorimeter (an area of size
$\Delta\eta\times\Delta\phi=0.125\times 0.175$ centred on the
barycenter of the photon cluster), as well as the small energy leakage
into the isolation cone, evaluated as functions of $\etg$ on simulated
samples of single photons, are subtracted from $\etiso$.
 
To match the definition between data and theory (\textit{i.e.}, fixed-order
pQCD calculations, which do not include pile-up or UE effects), a
correction to $\etiso$ is applied to account for the effects from the
UE and pile-up. This correction comes from the so-called ``jet-area''
method~\cite{jhep:0804:005,jhep:1004:065}. In this method, low-energy
jets are used to compute an ambient transverse energy density on an
event-by-event basis, which is then multiplied by the area of the
isolation cone and subtracted from the isolation transverse energy.
 
A data-driven correction is applied to the simulated
$\etiso$ variable to improve the agreement between MC and
data~\cite{JINST14:P12006}. The $\etiso$ distributions of
photon-enriched samples in data and MC
are fitted using a Crystal-Ball function and the correction is
computed as the difference in the fitted
mean between the two functions. After all these corrections to the
MC events, an improved description of the measured $\etiso$
distribution is obtained.
 
Corrections are also applied to the simulated events to match the
overall event conditions of the data sample and to account for known
differences between data and simulation. These additional corrections
include pile-up effects and photon identification, isolation and
reconstruction efficiency~\cite{JINST14:P12006}.
 
\textbf{Photon selection}.
The photon-candidate selection criteria applied are:
\begin{itemize}
\item The starting point of the selection is the photons
reconstructed and calibrated as described above. Both converted and
unconverted candidates are kept. Photons reconstructed near regions
of the calorimeter affected by read-out or high-voltage failures
are not considered.
\item The candidates are required to pass the tight identification
criterion described above.
\item Photons with $\etg>250$~\GeV\ and
$|\etag|<2.37$ are selected, excluding those in the transition
region ($1.37<|\etag|<1.56$) between the barrel and endcap calorimeters.
The threshold in $\etg$ at $250$~\GeV\ is chosen since this is the
region most sensitive to the proton PDFs. Isolated-photon production
cross sections at lower $\etg$ were measured by the ATLAS
Collaboration in previous
publications~\cite{pr:d83:052005,pl:b706:150,pr:d89:052004,jhep:1608:005,pl:b770:473,jhep:10:203}.
\item In events with multiple candidates satisfying these
requirements, the candidate with highest transverse energy
(leading photon) is retained for further study.
\item The $\etiso$ of the leading photon is required to be lower than
\etisocut. This requirement was optimised to retain most of the photons
satisfying the identification criteria, to obtain the best
signal-to-background ratio and to keep high and constant
the fraction of photon candidates that satisfy the isolation
selection on top of the identification
criteria~\cite{jhep:1608:005}. Two different samples are selected
using $R=0.4$ and $R=0.2$ for the radius of the isolation cone.
\end{itemize}
 
The number of data events selected by using the requirements listed
above amounts to $3\ 652\ 433$ for the $R=0.2$ sample and
$3\ 289\ 941$ for the $R=0.4$ sample. Each data sample is separated in
six $\etag$ regions to perform the cross section measurements
individually in each region, namely $|\etag|<0.6$, $0.6<|\etag|<0.8$,
$0.8<|\etag|<1.37$, $1.56<|\etag|<1.81$, $1.81<|\etag|<2.01$ and
$2.01<|\etag|<2.37$. The edges of these $\etag$ regions are driven by
the structure of the EM calorimeter. Each region in $|\etag|$ is divided into $12$
bins of $\etg$ with boundaries (in \GeV) set at $250$, $300$,
$350$, $400$, $470$, $550$, $650$, $750$, $900$, $1100$, $1500$,
$2000$ and $2500$. The binning is optimised according to the photon
energy resolution and the number of events per bin both in data and
MC. Some of the high-$\etg$ bins are not measured depending on the
$|\etag|$ region.
 
\section{Background evaluation and signal extraction}
\label{bacsub}
The main background to isolated-photon production arises from multi-jet
processes, in which a jet is misidentified as a photon. Such a jet
usually contains a light neutral meson, mainly a $\pi^0$, that carries
most of the energy of the jet and decays into two collimated photons.
A very small contribution from electrons or positrons misidentified as
photons is also present in the selected data samples.
 
\subsection{Multi-jet background}
For this study, a sample is obtained by applying all the selection
criteria described in Section~\ref{evesel}, except for the tight
identification and isolation requirements. Two subsamples are selected:
the subsample of
candidates that fulfill the requirements (tight subsample) and the
subsample of candidates that pass the loose criteria but fail some of
the tight requirements (non-tight subsample)~\cite{jhep:10:203}. The
non-tight subsample is expected to be enriched in background
candidates.
 
\begin{figure}[h]
\setlength{\unitlength}{1.0cm}
\begin{picture} (18.0,8.5)
\put (-3.9,-18.0){\includegraphics[width=23.1cm]{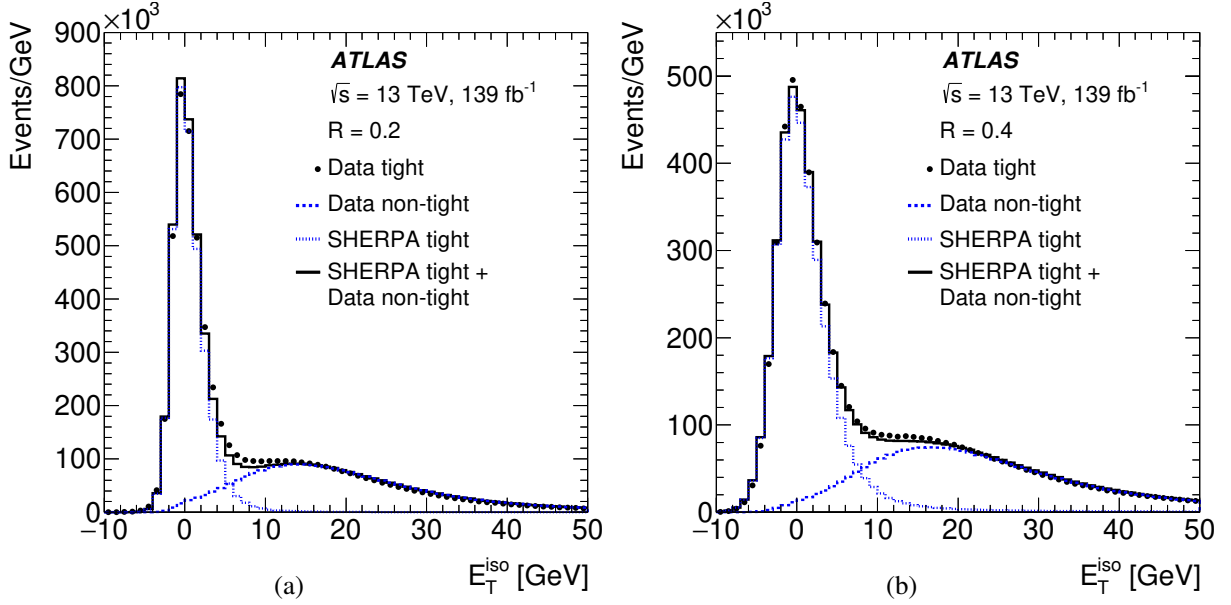}}
\end{picture}
\caption{
The $\etiso$ distributions with tight (dots) and non-tight (dashed
histograms, normalised according to the $\chi^2$ fit described in the text)
photon candidates in data with $\etg>250$~\GeV\ and $|\etag|<1.37$
or $1.56<|\etag|<2.37$ for $R=0.2$ (a) and $R=0.4$ (b). The
MC simulation of the signal using \sher\ is also shown (dotted
histogram, normalised according to the $\chi^2$ fit described in the
text). The solid histogram is the sum of the contributions of the
MC simulation of the signal using \sher\ and that of the non-tight
photon candidates and normalised according to the $\chi^2$ fit
described in the text.
}
\label{fig1}
\end{figure}

A clear signal peak of prompt photons can be observed in the $\etiso$
distribution of tight photon candidates in data as shown
in Figure~\ref{fig1}. In this figure, for illustrative purposes, the
result of a $\chi^2$ fit of the sum of the $\etiso$ templates from
\sher\ tight (signal) and data non-tight (background) photon
candidates to that of the tight photon data candidates is also
included. The signal and background components normalised
according to the fit are reported in the same figure.
The signal of prompt photons centred at
$\etiso=0$~\GeV\ is observed in both data and MC simulation. For the
$R=0.4$ tight data set, the signal peak around zero is wider and the tail at
high values of $\etiso$ is more populated than for the $R=0.2$ tight
data set. The non-tight $\etiso$ data distribution has a broad peak
around $\etiso\approx 15$~\GeV. This data set saturates the tail of
the distribution for larger $\etiso$ values and shows a tail towards
low values, which indicates the presence of background in the signal
region. A similar description of the data is obtained by using the
\pyt\ simulations for the signal instead of \sher. To avoid having to
rely on the $\etiso$ MC distribution for the signal, the multi-jet
background is subtracted using the data-driven method described
below.
 
The multi-jet background is subtracted using the same data-driven
method already employed in previous
publications~\cite{jhep:1608:005,pl:b770:473,jhep:10:203}. The
application of this method to the tight and non-tight subsamples is
briefly explained in the following. The multi-jet background
contamination is estimated and then subtracted by using a counting
technique based on the observed number of events in control regions of
the two-dimensional plane defined by using the photon identification
variable ($\gamma_{\mathrm{ID}}$) and the $\etiso$ variable. These two
variables are chosen because they are expected to be uncorrelated for
the background. In the following, the correlation correction factor
between the two variables in background events is denoted by
$\rbg$. The background subtraction is performed in each bin of $\etg$
separately for each $\etag$ region and each value of $R$.
 
Four regions are defined in the $\gamma_{\mathrm{ID}}-\etiso$ plane
based on the tight/non-tight $\gamma_{\mathrm{ID}}$ criteria and the
isolation ($\etiso<$ \etisocut) and non-isolation ($\etiso>$ \etisocutn)
requirements on the photon candidates. These four regions are defined as:
``A'' is the signal region, which contains tight and isolated photon candidates;
``B'' is the control region with non-isolated background events, which
contains tight and non-isolated photon candidates;
``C'' is the control region with non-tight background events, which
contains isolated and non-tight photon candidates;
``D'' is the control region that contains non-isolated
and non-tight photon candidates.
In addition, an upper limit on $\etiso$ of $50$~\GeV\ is also imposed
in regions B and D
to make the background subtraction less dependent on the MC
description of the data for higher $\etiso$ values.
These regions are defined
with a ``gap'' of 2 \GeV\ in $\etiso$ from region A, to have well
separated background-control and signal regions and minimise
migrations across the borders; the gap is chosen to be large enough in
comparison to any difference between data and simulations, while still
providing a sufficiently large number of events
in the control regions to perform the
data-driven subtraction. Other choices for the size of this gap and
for the upper limit in $\etiso$ are used to assess the corresponding
systematic uncertainties (see Section~\ref{choic}).
 
The relation between the number of signal events in region A
($N_A^{\mathrm{sig}}$) and the number of events in the control
regions is given by
 
\begin{equation}
N_A^{\mathrm{sig}}=N_A-\rbg\cdot(N_B-
f_B N_A^{\mathrm{sig}})\cdot
\frac{(N_C-
f_C N_A^{\mathrm{sig}})}{(N_D-
f_D N_A^{\mathrm{sig}})},
\label{eqone}
\end{equation}
where $N_K$ with $K=A,B,C,D$ is the number of observed events in each region
and
 
$$\rbg=\frac{N_A^{\mathrm{bg}}\cdot N_D^{\mathrm{bg}}}{N_B^{\mathrm{bg}}\cdot
N_C^{\mathrm{bg}}},$$
where $N_K^{\mathrm{bg}}$ with $K=A,B,C,D$
is the number of background events in each region; $\rbg$
is set to unity for the nominal results, the only
assumption in this method. This assumption is checked to be valid
within $(10-25)\%$, depending on the $\etg$ and $\etag$ region and
the isolation cone radius $R$. The differences of $\rbg$ with respect
to unity are included as systematic uncertainties in the final results
(see Section~\ref{syst_rbg}). Equation~(\ref{eqone}) takes
into account the expected number of signal events in the three
background control regions via the signal leakage fractions,
$f_K=N_K^{\mathrm{sig,MC}}/N_A^{\mathrm{sig,MC}}$ with $K=B,C,D$.
 
The signal leakage fractions are extracted from the MC simulations of
the signal, independently for each isolation-cone radius, using \sher\
and \pyt. Differences in the values of the signal
leakage fractions extracted from {\pyt} and {\sher} are observed. They
are due to the different treatment of the fragmentation component in
the two MC generators (see Section~\ref{datmc}).
 
\begin{figure}[p]
\setlength{\unitlength}{1.0cm}
\begin{picture} (18.0,19.0)
\put (-3.9,-7.5){\includegraphics[width=23.1cm]{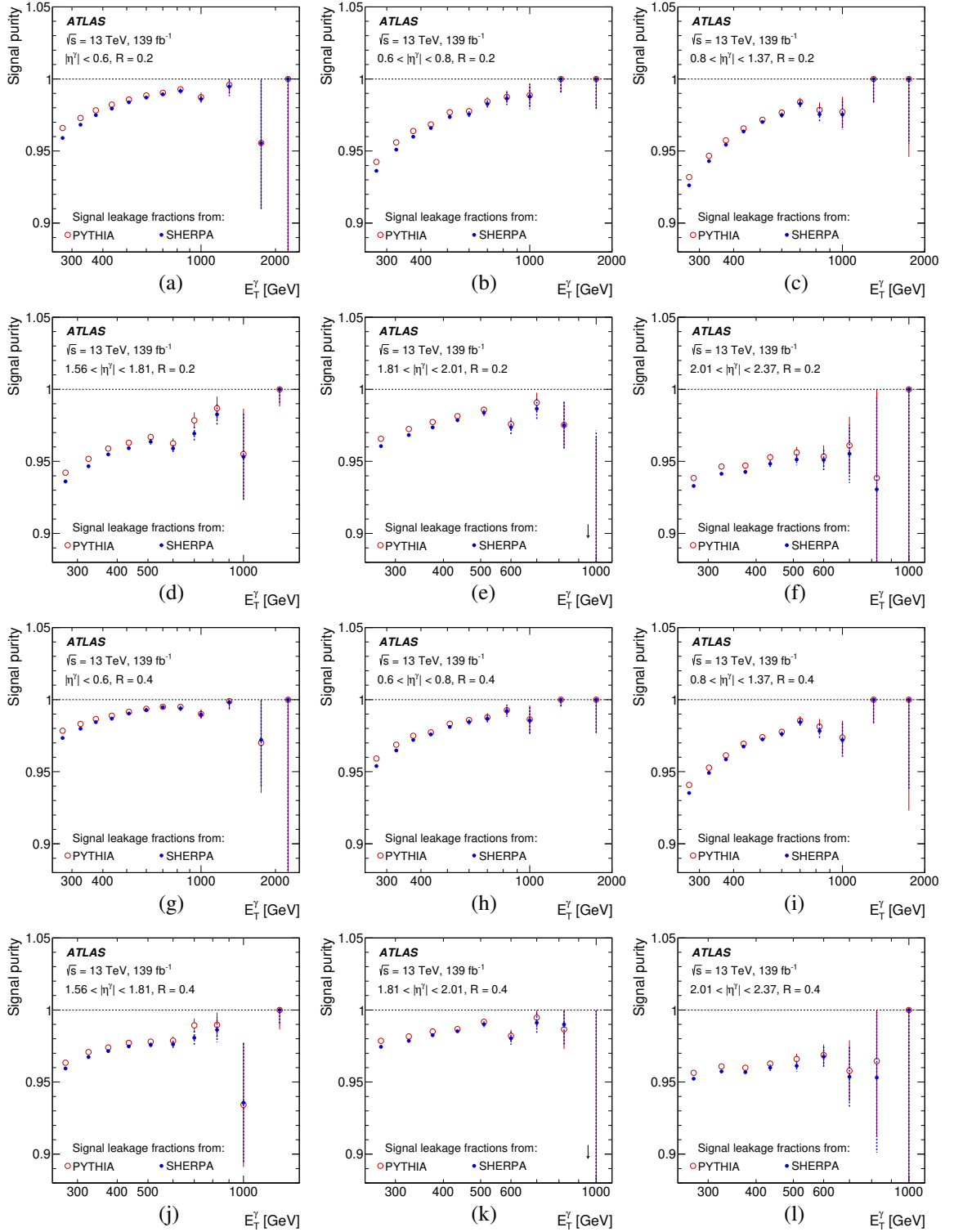}}
\end{picture}
\caption{
Estimated signal purities in data using the signal leakage fractions
from {\sher} (dots) and {\pyt} (open circles) as functions of
$\etg$ in different regions of $\etag$ for $R=0.2$ (a, b, c, d, e, f) and
$R=0.4$ (g, h, i, j, k, l). The data statistical uncertainties in the signal
purity are represented as solid (dashed) error bars for the
determination using the signal leakage fractions of \pyt\
(\sher). The arrows in (e) and (k) indicate the direction in which
the central values of the estimated signal purities are located
since they are outside of the plotted range in these bins.
}
\label{fig2}
\end{figure}

The signal yield is determined from the observed number of events in
the data in the four regions of the $\gamma_{\mathrm{ID}}-\etiso$
plane and the signal leakage fractions determined from the simulated
signal events using Equation~(\ref{eqone}). The signal purity, computed as
$P=N_A^{\mathrm{sig}}/N_A$, is shown in Figure~\ref{fig2} using the
signal leakage fractions from the {\sher} and {\pyt} signal samples.
The purity is $\gtrsim 90\%$ and very similar regardless of whether
\sher\ or \pyt\ samples are used to compute the signal leakage fractions. The
signal purity for $R=0.4$ is higher than for $R=0.2$. The nominal
signal yield is extracted using the signal leakage fractions from
\sher; the signal yield extracted from the signal leakage fractions of
\pyt\ is used to assess a systematic uncertainty in the purity
determination (see Section~\ref{sigmod}).
 
\FloatBarrier
 
\subsection{Background from electrons faking photons}
\label{backgepm}
Electrons and positrons can be misidentified as photons and they represent
an additional source of background. This background is largely
suppressed by the photon selection. The residual background
contribution is evaluated using the MC simulations from \sher~2.2.1
(see Section~\ref{datmc}) of the
$pp\rightarrow Z/\gamma^*\rightarrow e^+e^-$ and $pp\rightarrow
W\rightarrow e\nu$ processes. The electron background is
estimated separately in each $\etag$
region as a function of $\etg$ and found to be at a sub-percent level in the
phase-space region of this analysis, except for $1.81<|\etag|<2.37$
where it reaches $\sim 1\%$. The fraction of electrons faking photons
is found to be very similar for $R=0.2$ and $R=0.4$. Given the small
impact of this background, no attempt to subtract it is performed, and
a conservative systematic uncertainty equal to the size of the
evaluated background is assigned (see Section~\ref{noeb}).
 
\subsection{Signal yields}
\begin{figure}[h]
\setlength{\unitlength}{1.0cm}
\begin{picture} (18.0,20.0)
\put (-3.9,-6.5){\includegraphics[width=23.1cm]{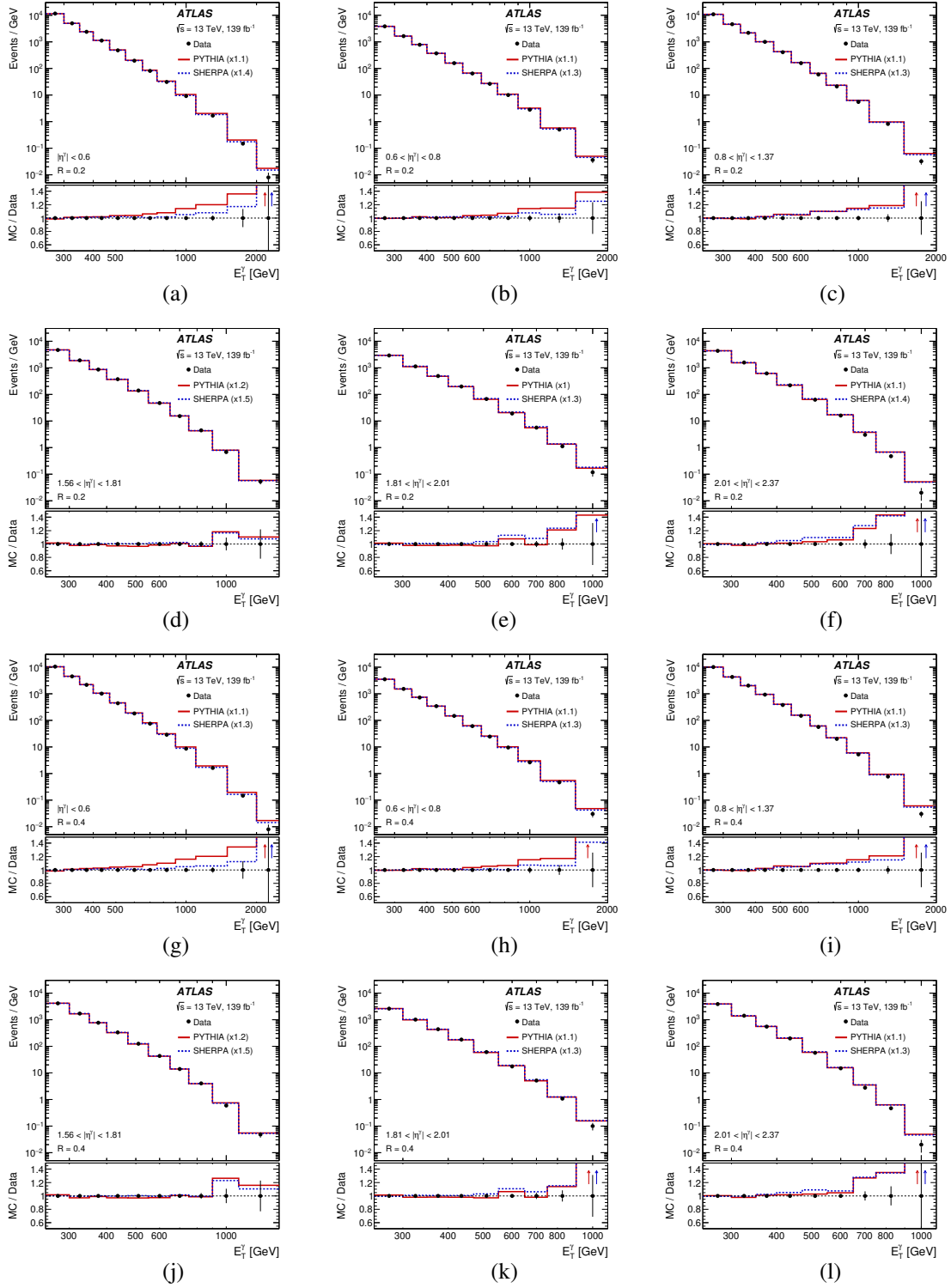}}
\end{picture}
\caption{
Estimated signal yields per \GeV\ in data (dots) using the signal leakage
fractions from {\sher} as functions of $\etg$ in different regions of
$\etag$ for $R=0.2$ (a, b, c, d, e, f) and $R=0.4$ (g, h, i, j, k,
l). For comparison, the MC simulations of the signal from {\sher}
(dashed histograms) and {\pyt} (solid histograms) are also
included. The MC distributions are normalised to the number of data
events in each $\etag$ region using the factors shown in
parenthesis. The ratio of the normalised MC and data distributions
is shown in the lower part of the figures. The error bars display
the statistical uncertainty of the data. The arrows in (a), (c), (e)
(f), (g), (h), (i), (k) and (l) indicate the direction in which
the ratio of the normalised MC and data distributions are located
since they are outside of the plotted range in these bins.
}
\label{fig3}
\end{figure}

The estimated signal yields using the signal leakage fractions from
{\sher} are shown in Figure~\ref{fig3} as functions of $\etg$ in
diffferent regions of $\etag$ for $R=0.2$ and $R=0.4$. The signal
yields using the signal leakage fractions from \pyt\ are very similar,
as evidenced by the similar signal purity (see Figure~\ref{fig2}). The
measured distributions decrease with increasing $\etg$ by
approximately six orders of magnitude within the measured range.
As expected, the signal yield for $R=0.2$ is larger than for
$R=0.4$. For comparison, the simulations of \pyt\ and \sher\ are also
included in these figures; both {\pyt} and {\sher} provide a
reasonable description of the shape of the data distribution within
statistical uncertainties, except at high $\etg$.
These predictions are based on tree-level calculations and, therefore,
are affected by a theoretical uncertainty due to missing higher-order
terms that can be as large as $50\%$.
 
\FloatBarrier
 
\section{Cross section measurement}
\label{fiducial}
The inclusive isolated-photon differential cross sections are measured
as functions of $\etg$ in the $\etag$ regions given by
$|\etag|<0.6$, $0.6<|\etag|<0.8$, $0.8<|\etag|<1.37$, $1.56<|\etag|<1.81$,
$1.81<|\etag|<2.01$ and $2.01<|\etag|<2.37$ for the two
isolation-cone radii, $R=0.2$ and $R=0.4$, separately.
The data are unfolded to particle level, as explained below, to the
region of fiducial phase space given by isolated photons with
$\etg>250$~\GeV\ and $|\etag|<2.37$, excluding the region
$1.37<|\etag|<1.56$.
The particle-level isolation ($\etisop$) on the photon is
built by summing the transverse energy of all stable particles, except
for muons and neutrinos, in a cone of radius $R=0.4$ or $R=0.2$ around
the photon direction, after the contribution from the UE is
subtracted; the same subtraction procedure used on data is applied at
the MC particle level. The particles associated with the overlaid $pp$
collisions are not considered in the calculation of the particle-level
isolation transverse energy; this is done to compare the measurements
to theoretical predictions without such an effect.
Isolation is ensured by requiring $\etisop<$ \etisocut.  The fiducial
phase-space region of the measurements follows closely the
detector-level event selection and it is indicated in
Table~\ref{tablezero}.
 
To study the dependence of isolated-photon production on the
isolation-cone radius, two additional measurements are performed, both
of which are based on the differential cross sections described
above. The first measurement is performed by integrating the
differential cross sections in each region of $\etag$ (`fiducial
integrated cross sections') and dividing by the width of each
$|\etag|$ region for each isolation-cone radius. These measurements
are sensitive to the dependence of the inclusive isolated-photon cross
section on $R$. The second measurement comprises the ratio of the
differential cross sections with $R=0.2$ and $R=0.4$ as a function of
$\etg$ in each $\etag$ region.
These ratios are performed using directly the measurements of the differential
cross sections for each isolation-cone radius.
In the evaluation of the statistical
uncertainties in data and MC simulations, the correlation between the
sample of photon candidates selected with $R=0.2$ and that with
$R=0.4$ is taken into account. Thanks to the cancellation of most of
the systematic uncertainties, this ratio provides a very stringent
test of the evolution of the $R$-dependence of the inclusive
isolated-photon differential cross section in $\etg$ for each $\etag$ region.
 
\subsection{Unfolding procedure for the
measurement of the differential   cross sections}
The data distributions, after background subtraction, as functions of
$\etg$ in the different $\etag$ regions defined above are unfolded to
the particle level, separately for $R=0.2$ and $R=0.4$. The unfolding
is performed independently for each value of $R$ and each $\etag$
region.
The iterative
application of Bayes' theorem is used to obtain the measured
differential cross sections. The Bayesian
unfolding~\cite{nim:a362:487} method as implemented in
RooUnfold~\cite{ijmpa:35:2050145} is used. In this method, the repeated
application of Bayes' theorem is used to invert the response
matrix. The response matrix is built from the two-dimensional
distribution in the $\etg$(reconstructed)--$\etg$(particle) plane of
the simulated events which fulfill simultaneously the full event
selection at reconstruction and particle levels; furthermore, in each
event the reconstructed photon is required to match the generated
photon within $\Delta R<0.2$. The two-dimensional distribution is then
used to calculate the probability for a photon generated with $\etg$(particle)
and $\etag$(particle) values to be reconstructed with $\etg$(reconstructed) and
$\etag$(reconstructed) values.
The method also accounts for the reconstructed photons which
are not matched to a truth photon because they are  outside of the
fiducial region (``reco unmatched'') as well as reconstruction
inefficiencies due to truth photons which are not matched to a
reconstructed photon (``truth unmatched''). The
regularisation parameter is the number of iterations
($N_{\mathrm{iter}}$), therefore regularisation is achieved by
stopping the iterative procedure at a given value of
$N_{\mathrm{iter}}$. The results are found to be fairly insensitive to
$N_{\mathrm{iter}}$; two iterations, \textit{i.e.} $N_{\mathrm{iter}}=2$, are
used in this analysis.
 
The nominal cross sections are measured using the response matrices
from the {\sher} samples and the deviations in the results obtained by using
{\pyt} instead are taken to represent systematic uncertainties of the
effect of the parton-shower and hadronisation models in the corrections
(see Section~\ref{unf}).
 
\begin{table}[h]
\caption{
Definition of the fiducial phase-space region for the measurements and
predictions.
}
\setlength{\unitlength}{1.0cm}
\begin{picture} (18.0,4.0)
\put (1.0,2.0){\scalebox{0.85}{\centerline{
\begin{tabular}{|l||c|c|c|c|c|c|} \hline
Requirement & \multicolumn{6}{c|}{Phase-space region} \\ \hline\hline
& \multicolumn{6}{c|}{ }
\\[-5pt]
$\etg$ & \multicolumn{6}{c|}{ $\etg>250$~\GeV }
\\[5pt] \hline
& \multicolumn{6}{c|}{ }
\\[-5pt]
Isolation & \multicolumn{6}{c|}{  $\etisop<$~\etisocut}
\\[5pt] \hline
&  &  & & & & \\[-5pt]
$\etag$ & $|\etag|<0.6$ & $0.6<|\etag|<0.8$ &
$0.8<|\etag|<1.37$ & $1.56<|\etag|<1.81$  & $1.81<|\etag|<2.01$ & $2.01<|\etag|<2.37$
\\[5pt] \hline
\end{tabular}
}}}
\end{picture}
\label{tablezero}
\end{table}
 
\FloatBarrier
 
\section{Systematic uncertainties}
\label{syst}
The sources of systematic uncertainties that affect the measurements
are the signal modelling, the background subtraction, the photon
reconstruction, the unfolding procedure, the running conditions and
the photon calibration. Each source is discussed in detail below.
For some of the systematic uncertainties, the Bootstrap
technique~\cite{as:7:1} is used to evaluate the
statistical uncertainty in the calculated values. The dependence of
the systematic uncertainties on $\etg$ is then fitted with smooth
functions using the estimated statistical uncertainties as inputs.
Each contribution to the systematic
uncertainty is assumed to be fully correlated between measurements
when calculating the uncertainties of the ratios of the
cross sections, except for the $\etiso$ modelling (see
Section~\ref{etiso_mod}). In the following text, an average value in
$\etag$ of the resulting uncertainty in the measured fiducial
integrated cross sections is quoted in parentheses for $R=0.2$ and
$R=0.4$, except in the cases for which the systematic uncertainty is
independent of $R$. The total systematic uncertainty and the main
contributions for the differential cross sections and the ratios are
discussed in Section~\ref{totsys}.
 
\subsection{Signal modelling}
\label{sigmod}
The uncertainty due to the signal modelling in the signal purity
calculation (see Section~\ref{bacsub}) is evaluated as the deviations
observed from the nominal result when using \pyt\ to compute the
signal leakage fractions.
The resulting
uncertainty in the measured cross sections is similar for both radii
($\pm 0.5\%$ for $R=0.2$ and $\pm 0.4\%$ for $R=0.4$).
 
\subsection{Background subtraction}
 
\subsubsection{Choice of background control regions}
\label{choic}
A data-driven method is used to subtract the multi-jet background in
the signal region. The estimation of the background contamination in
the signal region is affected by the choice of the background-enriched
control regions. For each modification of the background control
regions, the signal leakage fractions are recalculated.
 
\textbf{$\etiso$ requirement to define the control regions}. The
uncertainty due to the choice of the $\etiso$ requirement to define
the control regions is estimated by varying the $\etg$-dependent
isolation requirement from the nominal cut
($\etiso>\etisoc +2$~\GeV,
see Section~\ref{bacsub}) by $\pm 1$~\GeV. The resulting uncertainty
in the measured cross sections for $R=0.2$ is larger than for $R=0.4$
($\pm 0.05\%$ for $R=0.2$ and $\pm 0.01\%$ for $R=0.4$).
 
\textbf{Upper requirement on $\etiso$}. The dependence of the results on the
upper requirement on $\etiso$ for regions B and D is
estimated by removing it. Small differences are observed
on the resulting uncertainties in the measured cross sections between
$R=0.2$ and $R=0.4$ ($-0.06\%$ for $R=0.2$ and $-0.1\%$ for $R=0.4$).
 
\textbf{Identification criteria}.
The nominal
non-tight photon control region is defined by photons which pass
loose, but fail some of the tight identification criteria. The
uncertainty due to this choice is estimated by repeating the analysis
with three different non-tight definitions~\cite{pl:b770:473}. The
final uncertainty is estimated as the envelope of these variations. The
resulting uncertainty in the measured cross sections for $R=0.2$ is
somewhat larger than for $R=0.4$ ($\pm 0.8\%$ for $R=0.2$ and $\pm
0.6\%$ for $R=0.4$).
 
\subsubsection{Identification and isolation correlation in the
background}
\label{syst_rbg}
The isolation and identification photon variables used to define the
plane in the 2D side-band method to subtract the background (see
Section~\ref{bacsub}) are assumed to be uncorrelated for background
events ($\rbg=1$ in Equation~(\ref{eqone})). Any correlation between
these variables would affect the estimation of the signal purity
and lead to systematic uncertainties in the
background-subtraction procedure. The same data-driven method as used
in previous analyses~\cite{pl:b770:473,jhep:10:203} is applied for the
current analysis, using the same four validation regions.
Region $B$ is
subdivided into two regions:  region $B^{\prime}$ of tight photon
candidates with $\etisoc +2$~\GeV~$<\etiso < \etisoc+10$~\GeV\ and
region $B^{\prime\prime}$ of tight photon candidates with $\etiso >
\etisoc+10$~\GeV. Likewise, region $D$ is subdivided into two regions,
$D^{\prime}$ and $D^{\prime\prime}$, using the same separation in
$\etiso$ as above. The four regions $B^{\prime}$, $B^{\prime\prime}$,
$D^{\prime}$ and $D^{\prime\prime}$ are used to extract values of
$\rbg$ from the data after accounting for the signal
leakage fractions in those regions extracted either from \pyt\ or \sher\ MC
simulations. The dependence on the signal leakage is investigated by
increasing the lower limits on $\etiso$ for the validation regions,
$\etisoc+2$~\GeV\ ($\etisoc+10$~\GeV), each time by $1$~\GeV\ up to
$\etisoc+7$~\GeV\ ($\etisoc+15$~\GeV) for regions $B^{\prime}$ and
$D^{\prime}$ ($B^{\prime\prime}$ and $D^{\prime\prime}$), keeping the
width in $\etiso$ fixed to $8$~\GeV\ for the regions $B^{\prime}$ and
$D^{\prime}$.
As a result of this study,
the range of variation from unity for $\rbg$ is $0.10-0.25$. These
maximum deviations are used to re-evaluate the signal yields before
the unfolding procedure and are very similar for $R=0.2$ and $R=0.4$,
showing that  the effects are largely correlated.
The symmetrised resulting uncertainty in the measured cross
sections for $R=0.2$ is somewhat larger than for $R=0.4$ ($\pm 0.8\%$
for $R=0.2$ and $\pm 0.6\%$ for $R=0.4$).
 
\subsubsection{Background from electrons faking photons}
\label{noeb}
As discussed in Section~\ref{backgepm}, the background from electrons
faking photons is at a sub-percent level and no background subtraction is
performed. A systematic uncertainty is included by taking the full
size of this background, after adding $W+{\mathrm{jets}}$ and
$Z+{\mathrm{jets}}$ contributions linearly, depending on the $\etg$
and $\etag$ region. The resulting uncertainty in the measured cross
sections ranges from $\pm 0.4\%$ to $\pm 1.3\%$ for both
$R=0.2$ and $R=0.4$.
 
\subsection{Photon reconstruction}
 
\subsubsection{Photon-reconstruction efficiency}
The impact of the uncertainty in the photon-reconstruction efficiency is estimated
by propagating the uncertainties in the scale factors
applied to the MC events to match the reconstruction efficiency
between data and simulation (see
Section~\ref{evesel})~\cite{jhep:11:169} through the unfolding. The
resulting uncertainty in the measured cross sections is $\pm 0.3\%$
for both radii.
 
\subsubsection{Photon-identification efficiency}
The impact of the uncertainty in the photon-identification efficiency is estimated
by propagating the uncertainties in the scale factors, which are
applied to the MC events to match the tight identification efficiency
between data and simulation, to the final results. The resulting
uncertainty in the measured cross sections is $\pm 0.6\%$ for both
radii. The size of this systematic uncertainty is significantly
reduced~\cite{JINST14:P12006} with respect to the previous
analysis~\cite{jhep:10:203} ($1\%-3\%$).
 
\subsubsection{$\etiso$ modelling}
\label{etiso_mod}
The systematic uncertainty due to the modelling of the $\etiso$
distribution is obtained by propagating the uncertainties  in the
data-driven corrections to $\etiso$ applied to the MC samples
discussed in Section~\ref{evesel}. The resulting uncertainty in the
measured cross sections for $R=0.4$ is somewhat larger than for
$R=0.2$ ($\pm 0.02\%$ for $R=0.2$ and $\pm 0.07\%$ for $R=0.4$). This
source of uncertainty, in contrast to the others, is conservatively
taken as uncorrelated when performing the ratios
of the measured differential cross sections since the extraction procedures for
the two isolation radii are completely independent.
 
\subsection{Unfolding procedure}
\label{unf}
\subsubsection{Parton-shower and hadronisation model dependence}
The effect of the parton-shower and hadronisation models in the
unfolding is estimated as the change in the measured cross section
between the results using the response matrices of {\sher} (default MC
used for unfolding) and {\pyt}. Some differences are observed between
the resulting uncertainties in the measured cross sections for $R=0.2$
and $R=0.4$ ($\pm 0.7\%$ for $R=0.2$ and $\pm 0.5\%$ for $R=0.4$).
The uncertainties due to the scale variations, the strong coupling
constant and PDFs in the MC samples of events used for unfolding are
also investigated and found to have a negligible impact in the
measured cross sections.
 
\subsubsection{Unfolding closure}
An uncertainty due to the non-closure of the unfolding procedure is
estimated in the following way. The MC \sher\ distributions are
weighted to the data after background subtraction. The nominal MC
\sher\ samples are used as pseudo-data and unfolded with the weighted
samples. The unfolded results are compared to the \sher\ predictions
at particle level and the differences are taken as the non-closure uncertainties.
The resulting uncertainties in the measured
cross sections are typically much smaller than $0.1\%$, except in the
tails of the most forward $\etag$ region, where they reach up to
$0.3\%$, for both $R=0.2$ and $R=0.4$.
 
\subsubsection{MC statistical uncertainties}
The statistical uncertainty due to the limited number of simulated
events mainly affects the estimation of the response matrices. The
resulting uncertainties in the measured cross sections are very small
for both radii ($\pm 0.09\%$ for $R=0.2$ and $\pm 0.1\%$ for $R=0.4$).
 
\subsection{Running conditions}
\label{rc}
 
\subsubsection{Pile-up}
The uncertainty related to pile-up weighting of the simulated events
is propagated to the final results. The resulting
uncertainty in the measured cross sections for $R=0.4$ is larger than
for $R=0.2$ ($\pm 0.4\%$ for $R=0.2$ and $\pm 1.0\%$ for $R=0.4$).
 
\subsubsection{Trigger efficiency}
The uncertainty in the trigger efficiency is estimated using the same
methodology as in Ref.~\cite{epj:c77:317} and it is propagated to the
measured cross sections. The uncertainty is estimated to be between
$0.05\%$ and $0.15\%$, depending on the $\etag$ and $\etg$ regions and
independent of $R$.
 
\subsubsection{Measurement of the integrated
luminosity}
The uncertainty in the integrated luminosity is $\pm
1.7\%$~\cite{ATLAS-CONF-2019-021}.  This uncertainty is fully
correlated in all bins of all the measured cross sections.
 
\subsection{Photon calibration: energy scale and resolution}
\label{GES}
The assessment of the systematic uncertainty in the photon energy
scale and resolution is performed following the model originally
presented in Ref.~\cite{epj:c74:3071} and subsequently updated in
Ref.~\cite{JINST14:P12006} for Run 2 data-taking conditions.
 
The sources of uncertainty in the photon energy scale include: the
uncertainty in the overall energy scale adjustment using $Z\rightarrow
e^+e^-$ events; the uncertainty in the non-linearity of the energy
measurement at the cell level of the EM calorimeter; the uncertainty
in the relative calibration of the different calorimeter layers; the
uncertainty in the amount of material in front of the calorimeter; the
uncertainty in the modelling of the reconstruction of photon
conversions; and the uncertainty in the modelling of the lateral
shower shape. The sources of uncertainty in the photon energy
resolution include: the uncertainty in the modelling of the sampling
term and the uncertainty in the measurement of the constant term in
$Z$-boson decays. The sources of uncertainty are modelled using
independent components to account for their $\eta$ dependence. All the
uncertainty components are propagated separately through the analysis
to keep track of the information about the correlations between
different bins. The systematic uncertainty in the measured cross
section is evaluated by varying each individual source of uncertainty
separately by $\pm 1\sigma$ in the MC simulations and then adding the
uncertainty contributions in quadrature. The resulting uncertainties
in the measured integrated fiducial cross sections are $\pm 0.09\%$
for the energy resolution and $\pm 3.7\%$ for the energy scale,
independent of $R$. For the differential cross sections, the energy
scale uncertainty is $\approx (2-6)\%$ at $\etg=250$~\GeV\ and rises
up to $\approx (6-20)\%$ at high $\etg$, depending on the $\etag$
region, for both isolation-cone radii. This constitutes the dominant
contribution to the total systematic uncertainty (see Section~\ref{totsys}).
 
\subsection{Total systematic uncertainty}
\label{totsys}
\begin{figure}[h]
\setlength{\unitlength}{1.0cm}
\begin{picture} (18.0,19.0)
\put (-3.9,-7.5){\includegraphics[width=23.1cm]{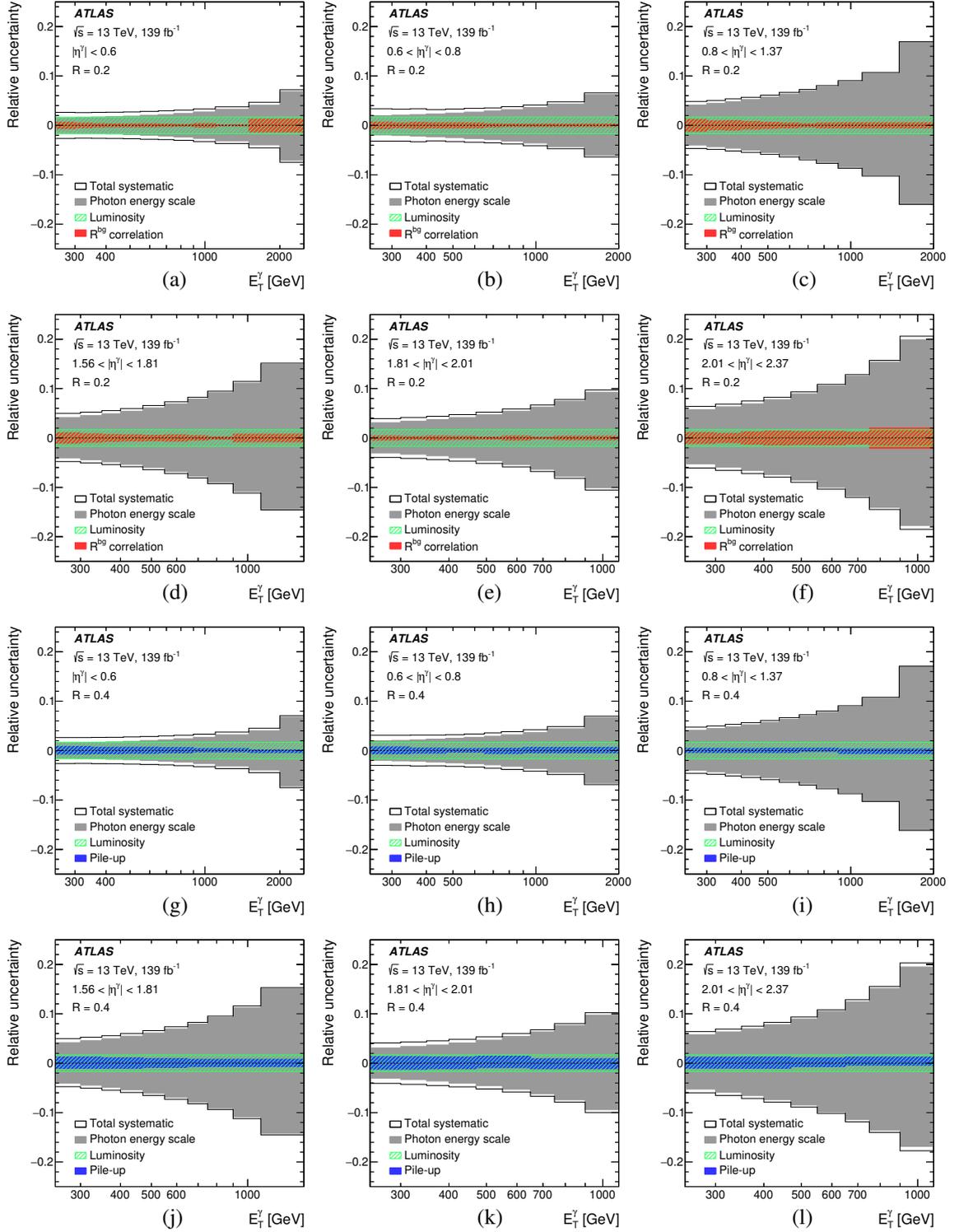}}
\end{picture}
\caption{
Relative systematic uncertainties in the differential cross
sections as functions of $\etg$ in different regions of $\etag$ for
$R=0.2$ (a, b, c, d, e, f) and $R=0.4$ (g, h, i, j, k, l): total
(black histograms), and main contributions from photon energy scale
(grey areas), luminosity (green hatched areas), $\rbg$ correlation
(red areas, only for $R=0.2$) and pile-up modelling (blue areas,
only for $R=0.4$).
}
\label{fig4}
\end{figure}

The total systematic uncertainty is computed by adding in quadrature
the sources of uncertainty listed in the previous sections.
Figure~\ref{fig4} shows the resulting relative total systematic
uncertainties in the differential cross sections as functions of $\etg$
in different regions of $\etag$ and for the two isolation radii.
There are bin-to-bin correlations of the systematic uncertainties for
each source. For instance, the systematic uncertainty due to the
photon energy scale and resolution is partially correlated bin-to-bin
and its decomposition into independent sources is used (see
Section~\ref{GES}). The following uncertainties are considered as
uncorrelated bin-to-bin: photon-identification efficiency, choice of
background control regions, $\etiso$ modelling and MC statistical
uncertainties.
 
The three dominant uncertainties in the measured differential cross sections,
namely, the photon energy scale and luminosity uncertainties for both
$R=0.2$ and $R=0.4$, and the uncertainty due to the background
correlation for $R=0.2$ and the pile-up uncertainty for $R=0.4$, are
also included in Figure~\ref{fig4}. The total systematic uncertainty
varies in the range $(3-20)\%$, depending on $\etg$ and $\etag$.
The systematic uncertainty in the photon-energy scale is larger in the
regions $0.8<|\etag|<1.37$ and $1.56<|\etag|<1.81$ due to the presence
of more material upstream of the calorimeter than in $|\etag|<0.8$. The
systematic uncertainties dominate the total uncertainty for $\etg$ up
to $1.5$~\TeV\ for $|\etag|<0.6$ and $0.8<|\etag|<1.37$, up to
$1.1$~\TeV\ for $0.6<|\etag|<0.8$ and $1.56<|\etag|<1.81$, and up to
$0.9$~\TeV\ for $1.81<|\etag|<2.37$. For higher $\etg$ values, the
statistical uncertainty of the data limits the precision of the
measurements. Previously~\cite{jhep:10:203}, the $\etg$ values up to
which the systematic uncertainties dominated were: $1.1$~\TeV\ for
$|\etag|<1.37$, $0.9$~\TeV\ for $1.56<|\etag|<1.81$, and $0.75$~\TeV\
for $1.81<|\etag|<2.37$.
 
\begin{figure}[h]
\setlength{\unitlength}{1.0cm}
\begin{picture} (18.0,10.0)
\put (-3.9,-16.0){\includegraphics[width=23.1cm]{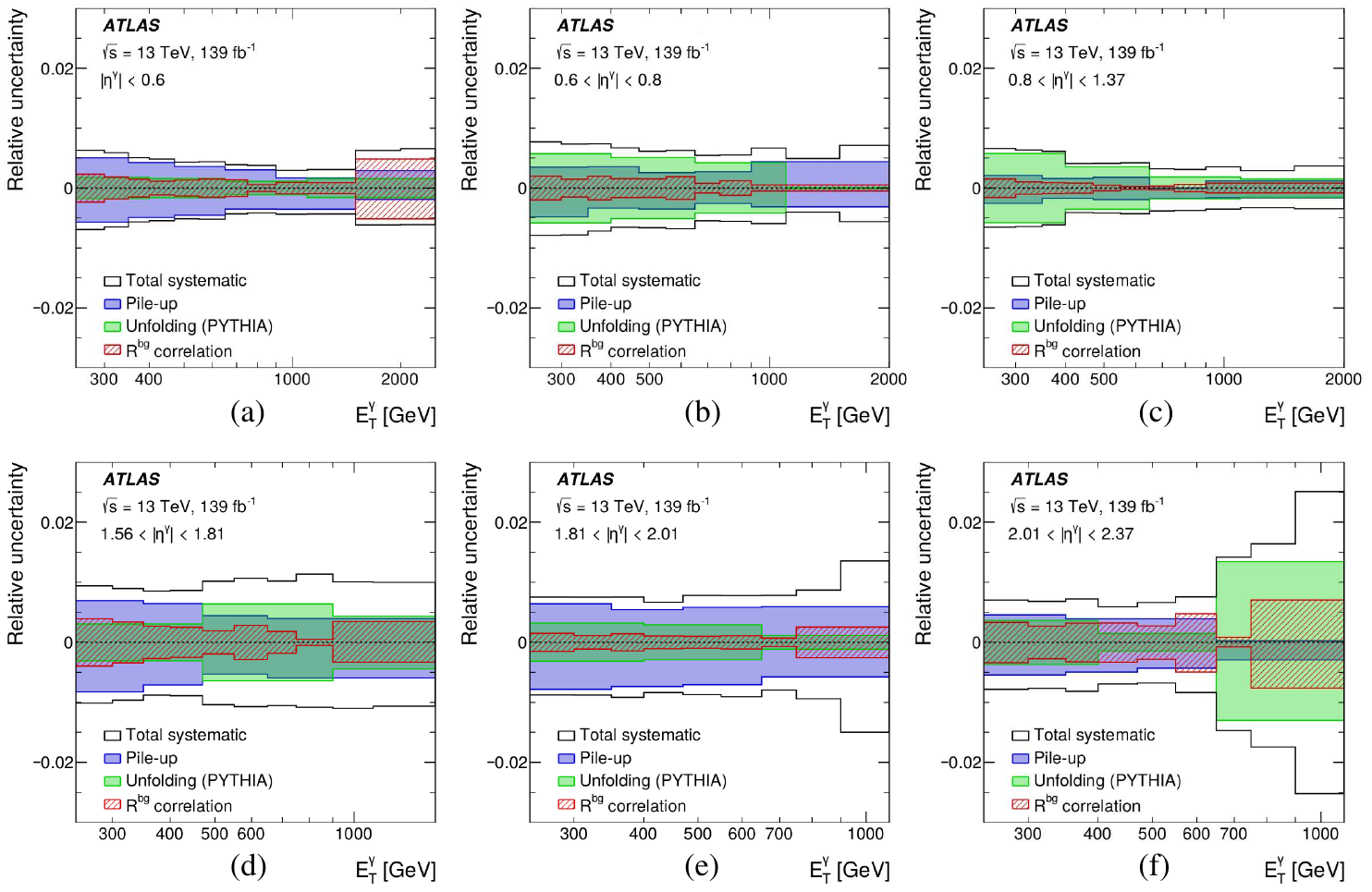}}
\end{picture}
\caption{
Relative total systematic uncertainty in the ratios of the
differential cross sections for $R=0.2$ and $R=0.4$ (black histograms),
relative uncertainty due to the pile-up modelling (blue areas),
relative uncertainty due to the MC modelling used for unfolding (green areas)
and relative uncertainty due to the $\rbg$ correlation (red hatched
areas) as functions of $\etg$ in different regions of $\etag$.
}
\label{fig5}
\end{figure}

The resulting relative total systematic uncertainties in the ratios of
the differential cross sections for $R=0.2$ and $R=0.4$ as functions
of $\etg$ in different regions of $\etag$ are shown in
Figure~\ref{fig5}.
Some residual statistical effects might remain in the individual
contributions to the total systematic uncertainty in the ratios after
taking into account the correlations between the measurements of the
differential cross sections with the different radii.
The main contributions to the total systematic
uncertainty in the ratios of the measured differential cross sections are
also included in Figure~\ref{fig5}; the dominant components
are the pile-up modelling, the MC modelling used for unfolding and the
$\rbg$ correlation. In these measurements, the luminosity and other
contributions which yield uncertainties in the differential cross
sections that are independent of the isolation
radius cancel out. In particular, the photon energy scale is no longer the
dominant contribution. Since the different sources of
uncertainty, except for the $\etiso$ modelling, are taken as fully
correlated, there is a significant reduction both in the total
systematic uncertainty (typically $<1\%$) and the data statistical
uncertainty. Thus, the ratios of the differential cross sections
constitute a compelling measurement for precise testing of the
underlying pQCD theory.
 
\FloatBarrier
 
\section{Theoretical predictions}
\label{nlo}
The NLO pQCD calculations presented in this paper are computed using the
programs {\jetp}~1.3.1\_2 and {\sher}~2.2.2. The NNLO pQCD predictions
are calculated in the \nnlojet\ framework. The comparison of these
predictions to the measurements is presented in Section~\ref{res}.
 
\textbf{\jetp\ predictions.} The \jetp\ program includes a full NLO
pQCD calculation of both the direct and the fragmentation contributions to
the cross section for the $\ppgj$ process. The number of massless
quark flavours is set to five. The renormalisation scale $\muR$, the
factorisation scale $\muF$ and the fragmentation scale $\muf$ are
chosen to be $\muR=\muF=\muf=\mu=\etg/2$. For the nominal predictions,
the calculations are performed using the MMHT2014~\cite{epj:c75:204}
PDF set and the BFG set II of parton-to-photon fragmentation
functions~\cite{epj:c2:529}, both at NLO. The strong coupling constant
is set to $\asz=0.120$; for consistency, in this calculation as well as in those
described below, the value of $\asz$ is set to that assumed in the PDF
set. For the electromagnetic coupling
($\alpha_{\textsc{em}}$), the low-energy limit value of $1/137.036$ is
used. The calculations are performed using the fixed-cone isolation
criterion at parton level which requires the total transverse energy
from the partons inside a cone of radius $R=0.4$ or $R=0.2$ around the
photon direction to be below \etisocut. Predictions based on other PDF
sets are also performed to test the sensitivity of the observables to
each different PDF set from the comparison to the data.
 
\textbf{\sher\ predictions.} The {\sher}~2.2.2 program consistently
combines parton-level calculations of $\gamma+ {\mathrm{(1,\ 2)-jet}}$
events at NLO and $\gamma+{\mathrm{(3,\ 4)-jet}}$ events at
LO~\cite{jhep:0202:044,prl:108:111601} supplemented with a parton
shower~\cite{jhep:0803:038} while avoiding double-counting
effects~\cite{jhep:1304:027}. A requirement on the photon isolation at
the matrix-element level is imposed using Frixione's
criterion
with ${\cal R}=0.1$, $n=2$ and $\epsilon=0.1$.
The prescription employed is referred to as  `hybrid-cone
isolation'~\cite{jp:g44:044007,jhep:04:166} since it includes the
application of the Frixione's criterion at a small value of $\Delta R$
(${\cal R}=0.1$) and the fixed-cone isolation at $R=0.4$ or $R=0.2$
used for the fiducial region of the measurement. Dynamic $\muR$ and
$\muF$ scales are adopted ($\muR=\muF=\etg$) as well as a dynamical merging
scale with $\bar{Q}_{\mathrm{cut}}=20$~\GeV~\cite{jp:g44:044007}. The
strong coupling constant is set to $\asz=0.118$. The same prescription
for the electromagnetic coupling as for the \jetp\ prediction is
used. Fragmentation into hadrons and simulation of the UE are
performed using the same models as for the LO \sher\ samples. The
NNPDF3.0 NNLO PDF set~\cite{jhep:1504:040} is used in conjunction with
the corresponding \sher\ tuning. These predictions are referred to as `\sher\
NLO' in the following.
 
\textbf{\nnlojet\ predictions.} The NNLO corrections include
three types of parton-level contributions, namely the two-loop
corrections to the Born-level processes, the one-loop Feynman diagrams
with an additional parton radiation, and the emission of two
additional partons. The three contributions to the NNLO corrections
are individually infrared divergent, but these divergencies cancel when
all contributions are considered together. Direct and
fragmentation processes are included in this calculation. The
fragmentation component is treated using the parton-to-photon
fragmentation functions BFG set II in the antenna approximation, as
described in Ref.~\cite{jhep:04:031}. Therefore, fixed-cone requirements,
as in the experiment, can be applied on these parton-level calculations.
The renormalisation and factorisation scales are set to
$\muR=\muF=\etg$, whereas the fragmentation scale is set to
$\muf=\sqrt{\etg\cdot E^{\mathrm{max}}_{\mathrm{T}}}\cdot
R$~\cite{jhep:08:094}, where $E^{\mathrm{max}}_{\mathrm{T}}$ is the
maximal hadronic transverse energy in the isolation cone of radius
$R$. The CT18NNLO PDF set~\cite{pr:d103:014013} is used. The strong
coupling constant is set
to $\asz=0.118$ while the electromagnetic coupling is set to
$\alpha_{\textsc{em}}=1/137.036$. The photon is required to be
isolated by demanding that the transverse energy within a cone of
$R=0.4$ or $R=0.2$ around the photon direction is smaller than
\etisocut. The prediction at NLO pQCD in the \nnlojet\ framework is
also calculated to illustrate the improvements achieved by including
the NNLO pQCD corrections.
 
\textbf{Differences between the theoretical calculations.} There are
several differences between the calculations using \jetp, \sher\ NLO
and \nnlojet: the calculations from \nnlojet\ include NNLO pQCD
corrections and adopt a different scheme to include the fragmentation
contribution as well as a different choice of $\muf$ than \jetp; the
calculations using \sher\ NLO include higher-order contributions as
well as parton showers. The application of the Frixione's criterion in
\sher\ NLO at matrix-element level allows the fragmentation
contribution to be ignored. The prediction for the cross section using
\sher\ NLO is at particle level and includes UE effects. A compilation
of the major features of the three different approaches is shown in
Table~\ref{tabletheory}.
 
\begin{table}[h]
\caption{
Major features of the three predictions used for inclusive
isolated-photon production.
}
\setlength{\unitlength}{1.0cm}
\begin{picture} (18.0,5.0)
\put (0.0,2.0){\scalebox{1.0}{\centerline{
\begin{tabular}{|l||l|l|l|l|l|l|} \hline
Program      & Order in $\as$ &  Fragmentation & Parton  &
Isolation  & PDF & Particle \\
&                        &                    &      shower &
method   &    & level \\
\hline\hline
\jetp\          & NLO                &
yes &  no &  fixed cone & $-$ MMHT2014 & no \\
& &    &     &        & $-$ CT18  &  \\
& &    &     &        & $-$ NNPDF3.1  &  \\
& &    &     &        & $-$ HERAPDF2.0  &  \\
& &    &     &        & $-$ ATLASpdf21  &  \\ \hline
\sher\ NLO & NLO for  $\gamma+ {\textrm{(1,\ 2)-jet}}$             &
no &  yes &  hybrid & NNPDF3.0 & yes \\
& LO for $\gamma+{\textrm{(3,\ 4)-jet}}$                 &       &        &       &    & \\ \hline
\nnlojet\     & (N)NLO             &
yes &  no &  fixed cone & CT18NNLO & no \\ \hline
\end{tabular}
}}}
\end{picture}
\label{tabletheory}
\end{table}
 
\begin{figure}[h]
\setlength{\unitlength}{1.0cm}
\begin{picture} (18.0,19.0)
\put (-3.9,-7.5){\includegraphics[width=23.1cm]{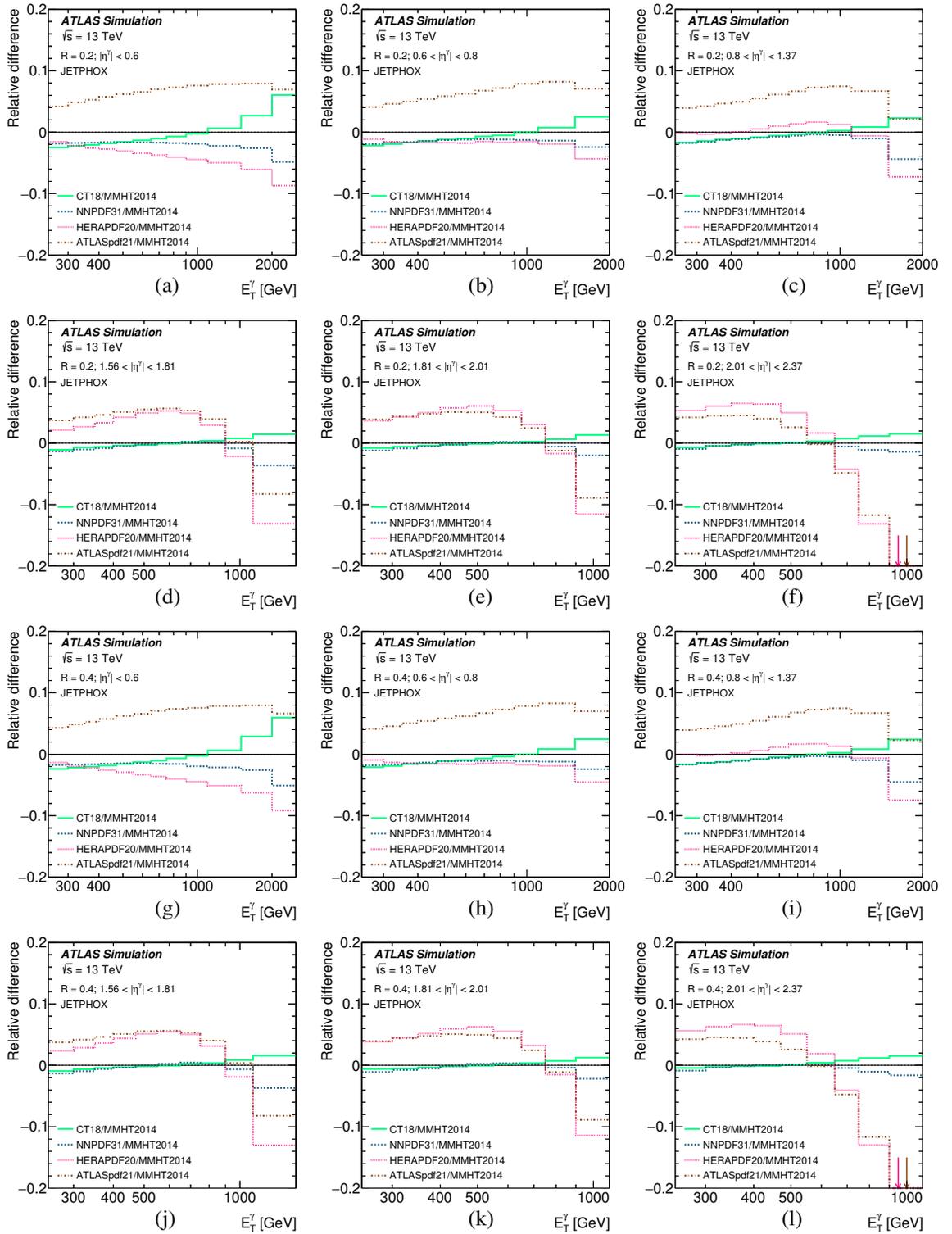}}
\end{picture}
\caption{
Relative difference between the \jetp\ predictions based on the CT18
(solid lines), NNPDF3.1 (dashed lines), HERAPDF2.0 (dotted lines)
and ATLASpdf21 (dot-dashed lines) and those based on the MMHT2014
PDFs for $R=0.2$ (a, b, c, d, e, f) and $R=0.4$ (g, h, i, j, k, l) as functions
of $\etg$ in different regions of $\etag$. The arrows in (f) and (l)
indicate the direction in which the relative differences are located
since they are outside of the plotted range in these bins.
}
\label{fig8}
\end{figure}

\textbf{Sensitivity of the NLO predictions to the PDFs.} The
sensitivity of the differential cross sections to the proton PDFs is
investigated by comparing the calculations of \jetp\ based on MMHT2014
with alternative calculations based on other PDF sets, namely
CT18NLO~\cite{pr:d103:014013}, NNPDF3.1~\cite{epj:c77:663},
HERAPDF2.0~\cite{epj:c75:580} and
ATLASpdf21~\cite{epj:c82:5}. The ATLASpdf21 PDFs are at NNLO, whereas
the other PDF sets are at NLO, and were extracted
including as input the ratios of the measured differential cross sections for
inclusive-photon production at 13~\TeV\ over those at
8~\TeV~\cite{jhep:04:093}.
Figure~\ref{fig8} shows the relative
difference between these alternative predictions of \jetp\ for $R=0.2$
or $R=0.4$ and the prediction based on MMHT2014 as functions
of $\etg$ in different regions of $\etag$. Differences between the
PDF sets are observed: the predictions based on the ATLASpdf21 and
on the HERAPDF2.0 PDF sets show differences of up to $\approx 10\%$
with respect to those based on the MMHT2014 PDF set, whereas those
from CT18NLO (NNPDF3.1) are within $2\%$ of those based on MMHT2014
at low $\etg$ but tend to be somewhat higher (lower) than MMHT2014
at high $\etg$ values in each $\etag$ region. As expected, the
sensitivity of the observables to the PDF set is largely independent
of the isolation cone radius.
 
\FloatBarrier
 
\subsection{Hadronisation and underlying-event corrections to the
fixed-order pQCD calculations}
\label{hadcor}
The NLO pQCD predictions from \jetp\ and the (N)NLO predictions from
\nnlojet\ are at the parton level, while the measurements are unfolded
at the particle level. Thus, there can be differences between the two
levels concerning the photon isolation as well as the photon
four-momentum. Since the data are corrected for pile-up and UE effects
and the distributions are unfolded to a phase-space definition in
which the requirement on $\etisop$ is applied after
subtraction of the UE, it is expected that the parton-to-hadron
corrections to the predictions are small.
 
Correction factors to the differential cross section predictions are
estimated by computing the ratio of the particle-level cross section
for a \pyt\ sample generated using version 8.243 with UE effects to the
parton-level cross section without UE effects. For the sample with UE,
the same jet-area based subtraction method used in data and
particle-level MC is applied. The correction factors are found to be
consistent with unity within $\pm 1\%$ for both isolation cone radii,
and no significant dependence on $\etag$ is observed. Thus, no
corrections are applied to the differential cross section predictions
and an uncertainty of $\pm 1\%$ is assigned for these effects for both
isolation radii.
 
For the ratio of the differential cross sections with different
isolation radii, the non-perturbative correction factors are also very
close to unity; a fit to a constant function for the non-perturbative
correction for the ratio of cross sections yields $0.9998\pm
0.0008$. Also in this case, no correction is applied and an
uncertainty is assigned to the ratio predictions given by the
difference of the correction factors from unity in each bin of
$\etg$. This uncertainty ranges from $0.06\%$ to $0.8\%$, depending on
the $\etg$ bin and the $\etag$ region.
 
\subsection{Theoretical uncertainties}
\label{thunc}
The theoretical uncertainties for the differential cross section
predictions are estimated in the following way:
 
\begin{itemize}
\item The uncertainty in the NLO pQCD predictions from \jetp\ due to
missing higher-order terms is estimated by repeating the calculations using values of
$\muR$, $\muF$ and $\muf$ scaled by the factors $0.5$ and $2$. The
three scales are either varied simultaneously, individually or by
fixing one and varying the other two. In all cases, the condition
$0.5\leq\mu_A/\mu_B\leq 2$ is imposed, where $A,B={\mathrm{R,\ F,\
f}}$. The final uncertainty is taken as the largest deviation from
the nominal value among the $14$ possible variations. A similar method
is used for the predictions of \nnlojet~\cite{jhep:08:094}. In the case
of the \sher\ NLO predictions, which do not include the fragmentation
contribution, $\muR$ and $\muF$ are varied as above and the largest
deviation from the nominal value among the $6$ possible variations is
taken as the uncertainty.
\item The uncertainty in the NLO pQCD predictions from \jetp\ due to the
uncertainty in the proton PDFs is estimated by repeating the
calculations using the $50$ sets from the MMHT2014 error
analysis~\cite{epj:c75:204} and applying the Hessian
method~\cite{pr:d65:014013} for the evaluation of the PDF
uncertainty. The PDF uncertainty for the \nnlojet\ calculations is not
available; thus, this uncertainty is taken from the corresponding relative
uncertainty of the \jetp\ predictions. In the case of \sher\ NLO, this
uncertainty is estimated using the $100$ replicas from the NNPDF3.0
analysis~\cite{jhep:1504:040}.
\item The uncertainty in the NLO pQCD predictions from \jetp\ (\sher\ NLO)
due to the uncertainty in $\as$ is estimated by repeating the
calculations using two additional sets of proton PDFs from the
MMHT2014 (NNPDF3.0) analysis, for which different values of $\as$ at
$\mz$ are assumed in the fits, namely $0.118$ ($0.117$) and $0.122$
($0.119$); in this way, the correlation between $\as$ and the PDFs is
preserved. The $\as$ uncertainty for the \nnlojet\ calculations is not
available; thus, this uncertainty is taken from the corresponding relative
uncertainty of the \jetp\ predictions.
\item The uncertainty in the NLO pQCD predictions from \jetp\ due to
the uncertainty in the fragmentation functions is evaluated by
repeating the calculations using the BFG set I~\cite{epj:c2:529} and
comparing the results with the nominal predictions. The uncertainty is
found to be negligible.
\item An uncertainty of $\pm 1\%$ is included in the
uncertainty of the \jetp\ and \nnlojet\ predictions due to the non-perturbative
corrections.
\end{itemize}
 
\begin{figure}[h]
\setlength{\unitlength}{1.0cm}
\begin{picture} (18.0,19.0)
\put (-3.9,-7.5){\includegraphics[width=23.1cm]{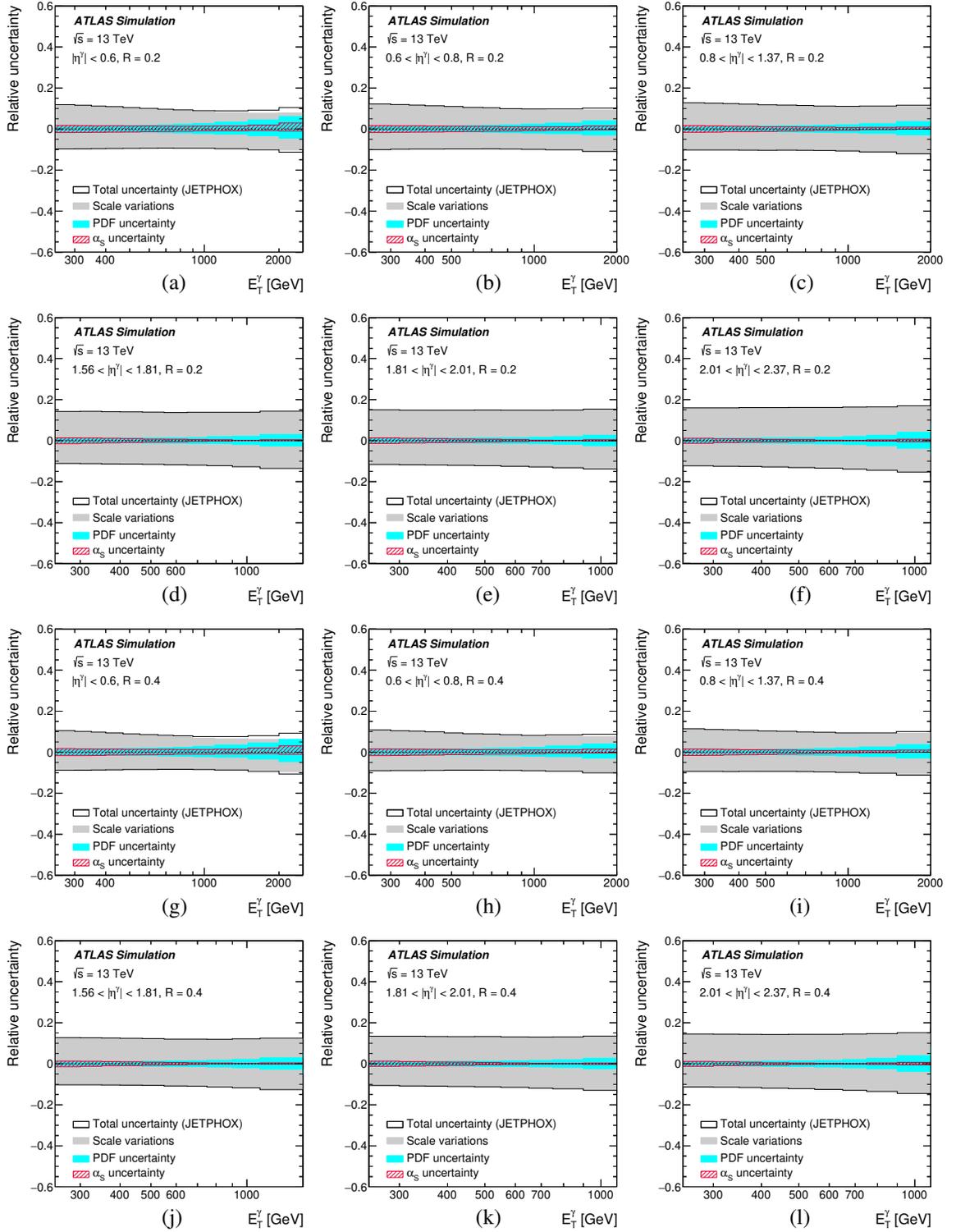}}
\end{picture}
\caption{
Relative theoretical uncertainty in \jetp\ arising from
scale variations (grey areas), PDF uncertainty (cyan areas), $\as$
uncertainty (red hatched areas) and the total theoretical uncertainty (black
histogram, which includes the uncertainty in the non-perturbative
corrections) for the differential cross sections with $R=0.2$ (a, b,
c, d, e, f) and $R=0.4$ (g, h, i, j, k, l) as functions of $\etg$ in
different regions of $\etag$.
}
\label{fig9}
\end{figure}

\begin{figure}[h]
\setlength{\unitlength}{1.0cm}
\begin{picture} (18.0,19.0)
\put (-3.9,-7.5){\includegraphics[width=23.1cm]{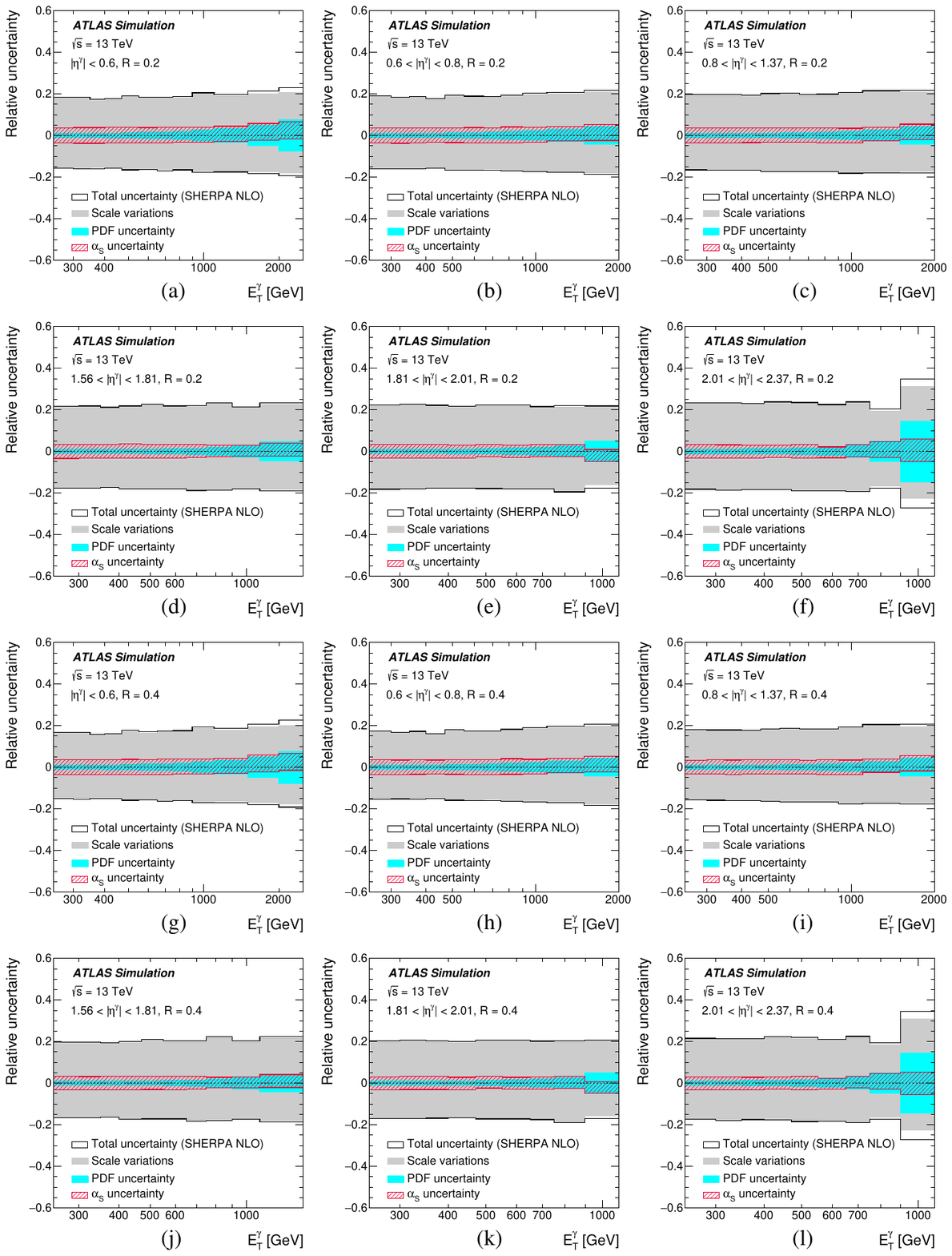}}
\end{picture}
\caption{
Relative theoretical uncertainty in \sher\ NLO arising from
scale variations (grey areas), PDF uncertainty (cyan areas), $\as$
uncertainty (red hatched areas) and the total theoretical uncertainty (black
histogram) for the differential cross sections with $R=0.2$ (a, b,
c, d, e, f) and $R=0.4$ (g, h, i, j, k, l) as functions of $\etg$ in
different regions of $\etag$.
}
\label{fig10}
\end{figure}

\begin{figure}[h]
\setlength{\unitlength}{1.0cm}
\begin{picture} (18.0,19.0)
\put (-3.9,-7.5){\includegraphics[width=23.1cm]{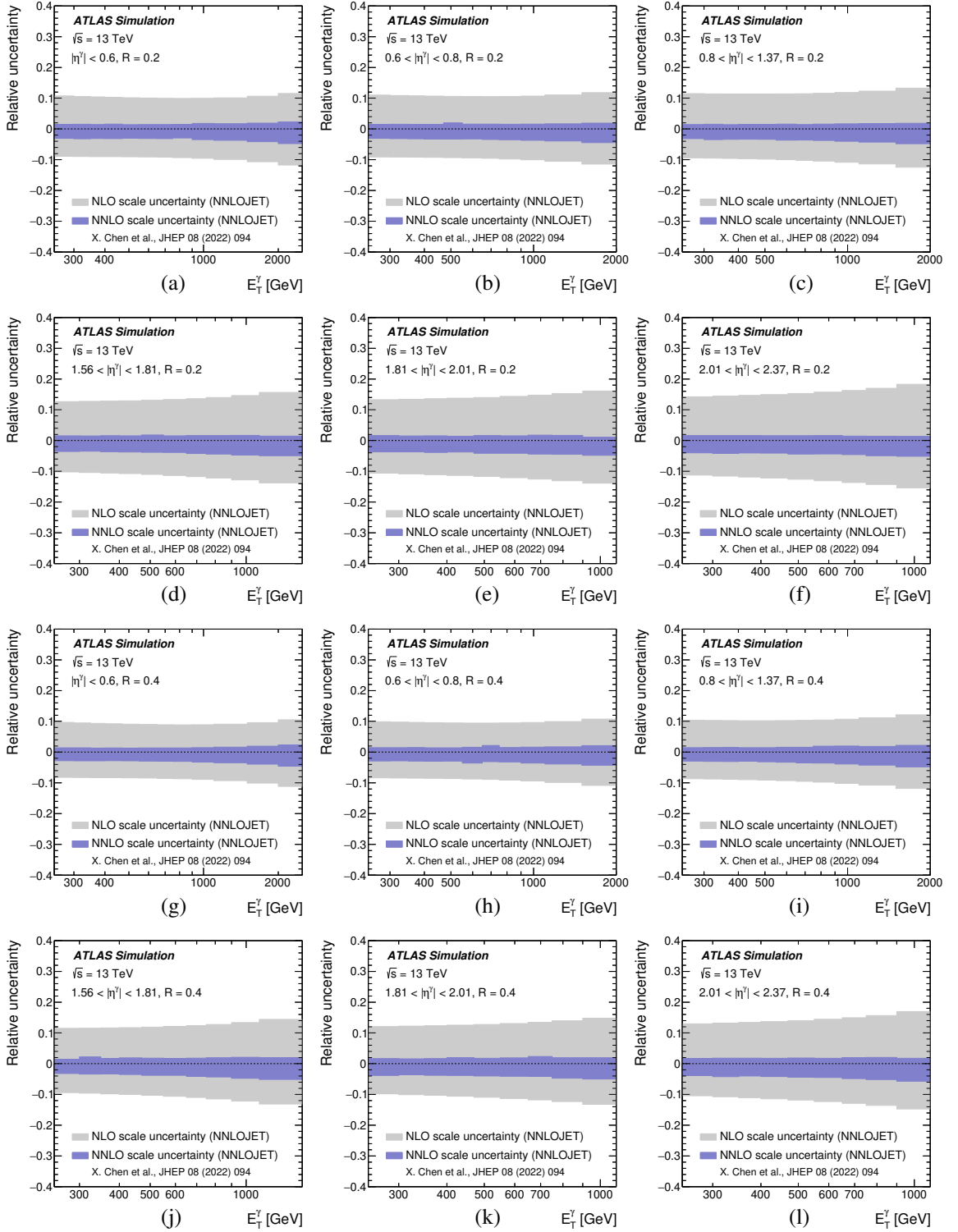}}
\end{picture}
\caption{
Relative theoretical uncertainty in \nnlojet\ arising from
scale variations in the NLO (grey areas) and NNLO (violet areas)
predictions for the differential cross sections with $R=0.2$ (a, b,
c, d, e, f) and $R=0.4$ (g, h, i, j, k, l) as functions of $\etg$ in
different regions of $\etag$.
}
\label{fig10_nnlojet}
\end{figure}

The total theoretical uncertainty for the differential cross section
predictions is obtained by adding in quadrature the individual
uncertainties listed above. Figures~\ref{fig9} and \ref{fig10} show
the relative total theoretical uncertainties and the components as functions
of $\etg$ in different regions of $\etag$ for $R=0.2$ and $R=0.4$ for
\jetp\ and \sher\ NLO, respectively. The total theoretical uncertainty
ranges from $\approx 10\%$ to $\approx 15\%$ for \jetp\ and it is
$\approx 20\%$ for \sher\ NLO. No significant difference in the size
of the uncertainties is observed between the predictions for
$R=0.2$ and $R=0.4$. The dominant theoretical uncertainty is
the one arising from missing higher-order terms. For large $\etg$ values,
the uncertainty coming from the PDFs is the second dominant
contribution. Figure~\ref{fig10_nnlojet} shows the uncertainties in
the NNLO pQCD prediction due to missing higher-order terms. These
uncertainties are in the range $(1-6)\%$ and are smaller than those in
the NLO pQCD prediction (also shown in Figure~\ref{fig10_nnlojet}) by a
factor between $2-15$, depending on $\etg$, $\etag$ and $R$.
 
\begin{figure}[h]
\setlength{\unitlength}{1.0cm}
\begin{picture} (18.0,19.0)
\put (-3.9,-7.5){\includegraphics[width=23.1cm]{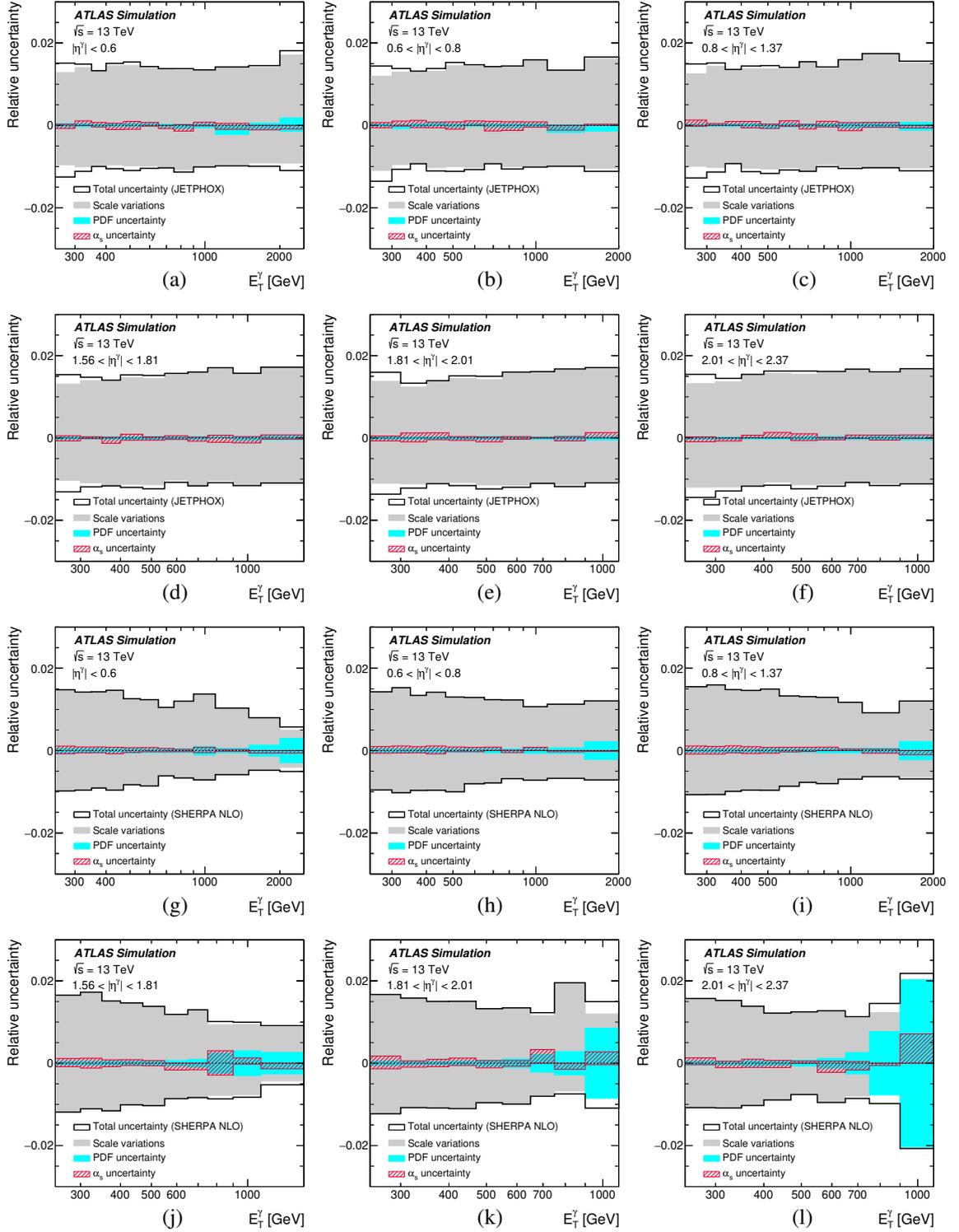}}
\end{picture}
\caption{
Relative theoretical uncertainty in \jetp\ (a, b, c, d, e, f) and \sher\ NLO (g, h, i, j, k, l)
arising from scale variations (grey areas), PDF uncertainty (cyan
areas), $\as$ uncertainty (red hatched areas) and the total theoretical
uncertainty (black histogram, which includes the uncertainty in the
non-perturbative corrections in the case of \jetp) for the ratio of the
differential cross sections as functions of $\etg$ in different regions of $\etag$.
}
\label{fig11}
\end{figure}

\begin{figure}[h]
\setlength{\unitlength}{1.0cm}
\begin{picture} (18.0,8.0)
\put (-3.9,-18.0){\includegraphics[width=23.1cm]{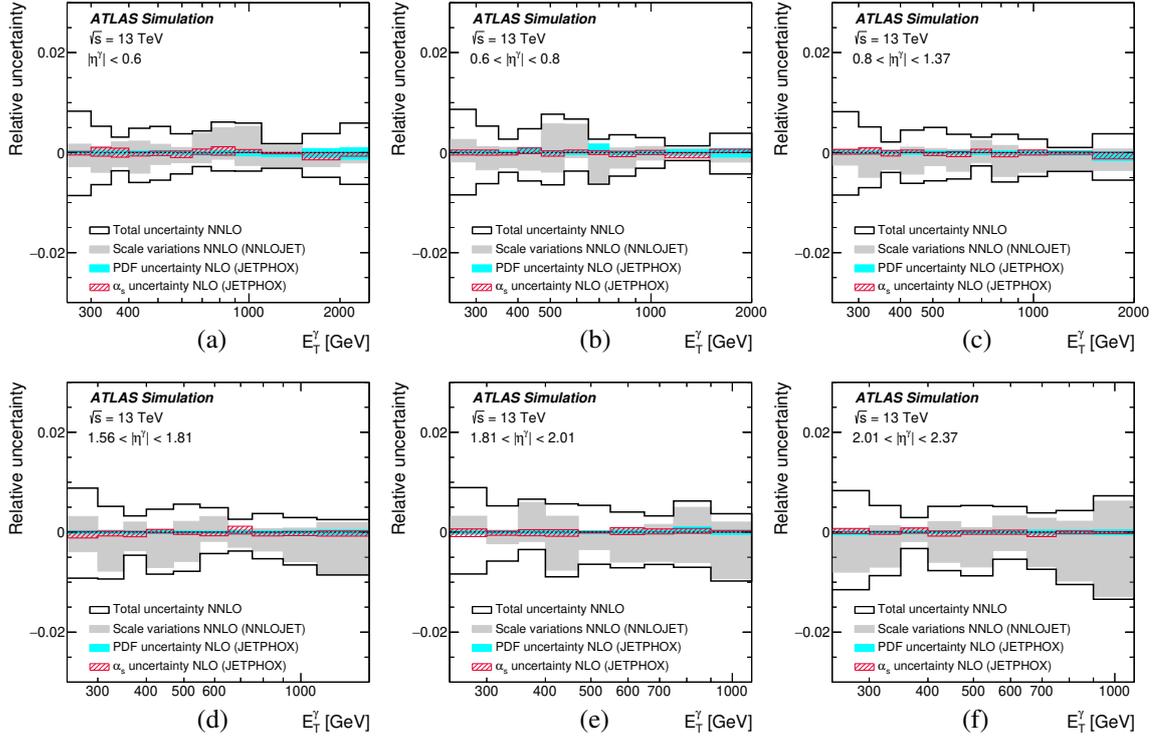}}
\end{picture}
\caption{
Relative theoretical uncertainty in the NNLO pQCD prediction for the
ratio of the differential cross sections as functions of $\etg$ in
different regions of $\etag$ from \nnlojet\ due to scale variations
(grey areas). The relative theoretical uncertainty in the
NLO pQCD prediction for the ratio from \jetp\ due to the uncertainty
in the PDFs (cyan areas) and the uncertainty in $\as$ (red hatched areas)
are also shown. The total relative theoretical uncertainty in the
ratio is shown as the black histogram and also includes the
uncertainty in the non-perturbative corrections and the statistical
uncertainty in the NNLO pQCD predictions.
}
\label{fig11_nnlojet}
\end{figure}

Figure~\ref{fig11} shows the relative theoretical uncertainty and its
components as functions of $\etg$ in different regions of $\etag$ for
the ratio of the differential cross sections for the \jetp\ and \sher\
NLO predictions. The theoretical uncertainties in the ratios are estimated as
fully correlated for both isolation-cone radii; as a consequence, a
significant reduction of the theoretical uncertainty is obtained: for \jetp\
(\sher\ NLO), the theoretical uncertainty decreases from $\approx (10-15)\%$
($\approx 20\%$) in the differential cross sections to $\approx 1.5\%$
($\approx 1.5\%$) in the ratios. Figure~\ref{fig11_nnlojet} shows the
relative theoretical uncertainty and its components as functions
of $\etg$ in different regions of $\etag$ for the ratio of the
differential cross sections for the NNLO predictions; the uncertainty
due to missing higher-order terms decreases to typically less than
$1\%$ in the ratio.
 
\FloatBarrier
 
\clearpage
 
\section{Results}
\label{res}
\subsection{Differential cross sections as functions
of $\etg$ in different $\etag$ regions}
 
\begin{figure}[h]
\setlength{\unitlength}{1.0cm}
\begin{picture} (18.0,10.0)
\put (-3.9,-16.0){\includegraphics[width=23.1cm]{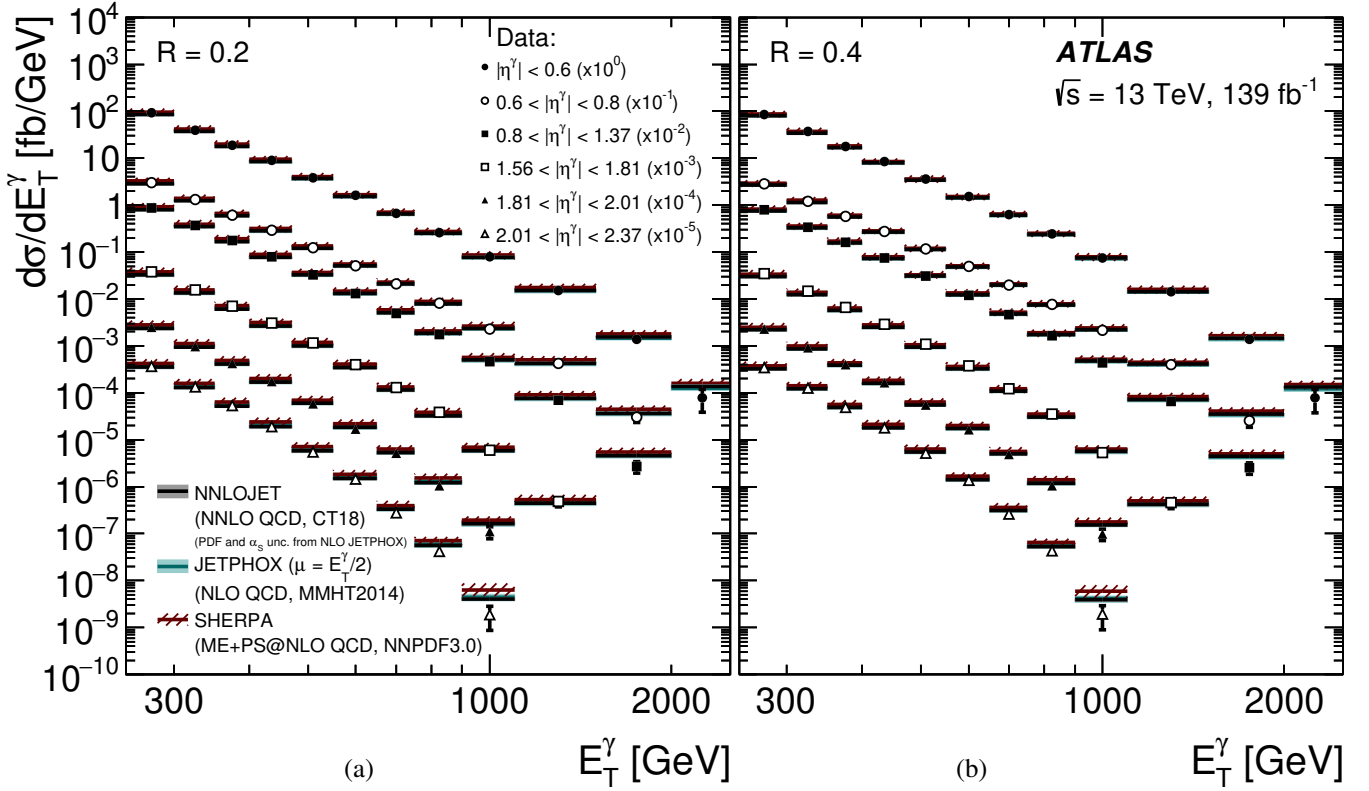}}
\end{picture}
\caption{
Measured differential cross sections for inclusive isolated-photon
production as functions of $\etg$ in $|\etag|<0.6$ (dots),
$0.6<|\etag|<0.8$ (open circles), $0.8<|\etag|<1.37$ (black
squares), $1.56<|\etag|<1.81$ (open squares), $1.81<|\etag|<2.01$
(black triangles) and $2.01<|\etag|<2.37$ (open triangles) for
$R=0.2$ (a) and $R=0.4$ (b). The NLO pQCD predictions from
\jetp\ (blue lines) based on the MMHT2014 PDFs, the ME+PS@NLO QCD
predictions from \sher\ NLO (brown lines) based on the NNPDF3.0 PDFs
and the NNLO pQCD predictions from \nnlojet\ based on the CT18NNLO
PDFs (black lines) are also shown.
The measurements and the predictions are normalised by the factors
shown in parentheses for each $\etag$ region to aid visibility. The
error bars represent the statistical and systematic uncertainties added in
quadrature. For most of the points, the error bars are smaller than
the marker size and, thus, not visible. The hatched and shaded bands
represent the theoretical uncertainty.
}
\label{fig12}
\end{figure}

Figure~\ref{fig12} shows the inclusive isolated-photon differential
cross sections as functions of $\etg$ in different regions of $\etag$
for $R=0.2$ and $R=0.4$. The measured cross sections decrease by
approximately six orders of magnitude in the investigated range.
The shape of the measured cross sections is similar for different
$\etag$ regions and radii, though the normalisation of the
measurements for $R=0.2$ is higher than for $R=0.4$. Values of
$\etg$ up to $2.5$~\TeV\ are measured with the full Run 2 ATLAS data set.
 
The NLO pQCD predictions of \sher\ NLO and \jetp\ and the NNLO pQCD
predictions of \nnlojet\ are compared to the measurements in
Figure~\ref{fig12}. These predictions are consistent with each other
within the theoretical uncertainties.
 
\begin{figure}[p]
\setlength{\unitlength}{1.0cm}
\begin{picture} (18.0,22.0)
\put (-3.9,-3.5){\includegraphics[width=23.1cm]{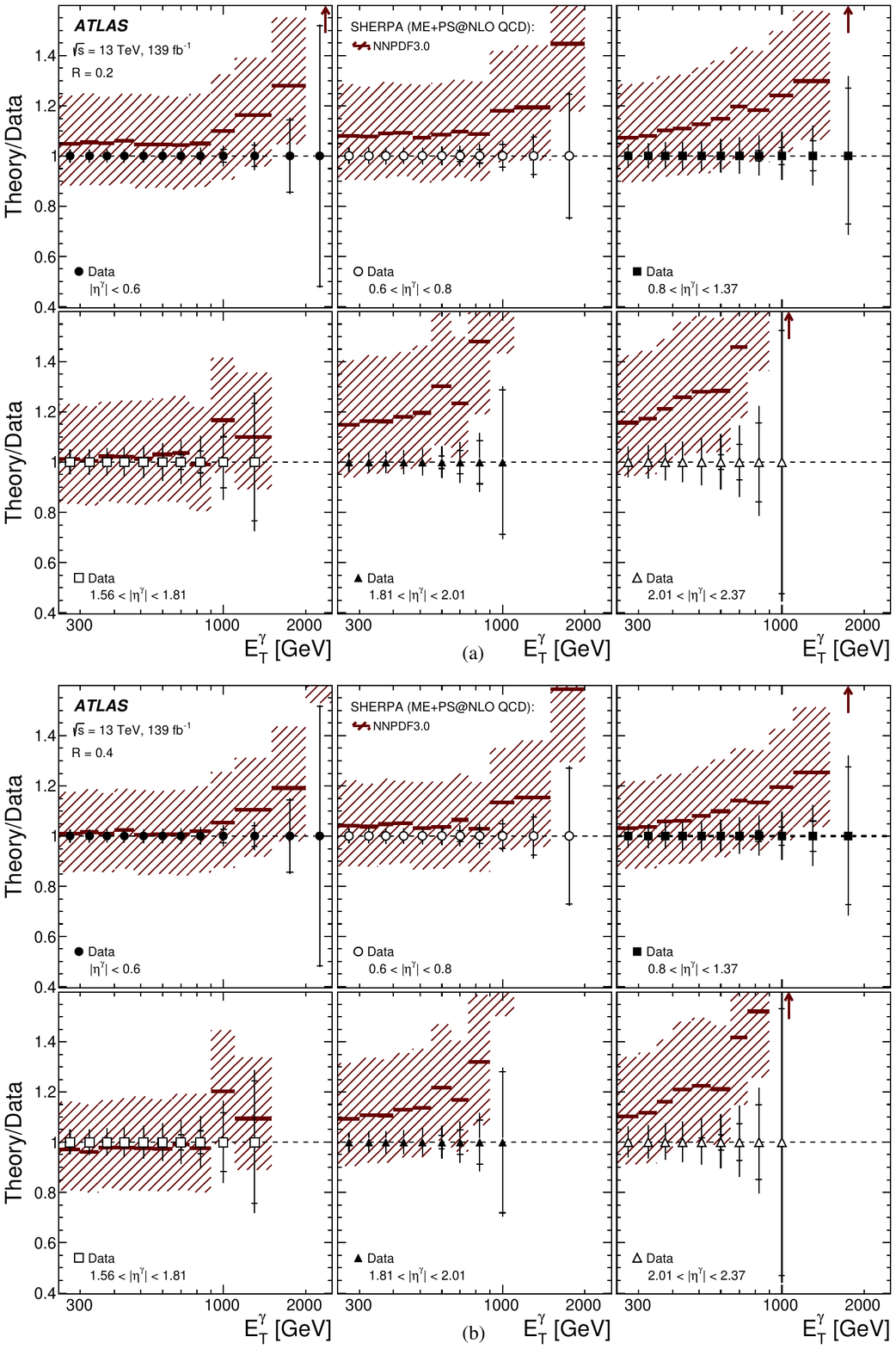}}
\end{picture}
\caption{
Ratio of the NLO pQCD calculations from \sher\ NLO based on the NNPDF3.0 PDF
set and the measured differential cross sections for
inclusive isolated-photon production with $R=0.2$ (a) and
$R=0.4$ (b) as functions of $\etg$ in different regions of $\etag$.
The inner (outer) error bars represent the statistical uncertainties
(statistical and systematic uncertainties added in quadrature). For
most of the points, the inner error bars are smaller than the marker size
and, thus, not visible. The hatched bands represent the theoretical
uncertainty. The arrows indicate the direction in which the ratios of
the calculations from \sher\ NLO and the measured differential cross
sections are located since they are outside of the plotted range in
these bins.
}
\label{fig13}
\end{figure}

\begin{figure}[p]
\setlength{\unitlength}{1.0cm}
\begin{picture} (18.0,22.0)
\put (-3.9,-3.5){\includegraphics[width=23.1cm]{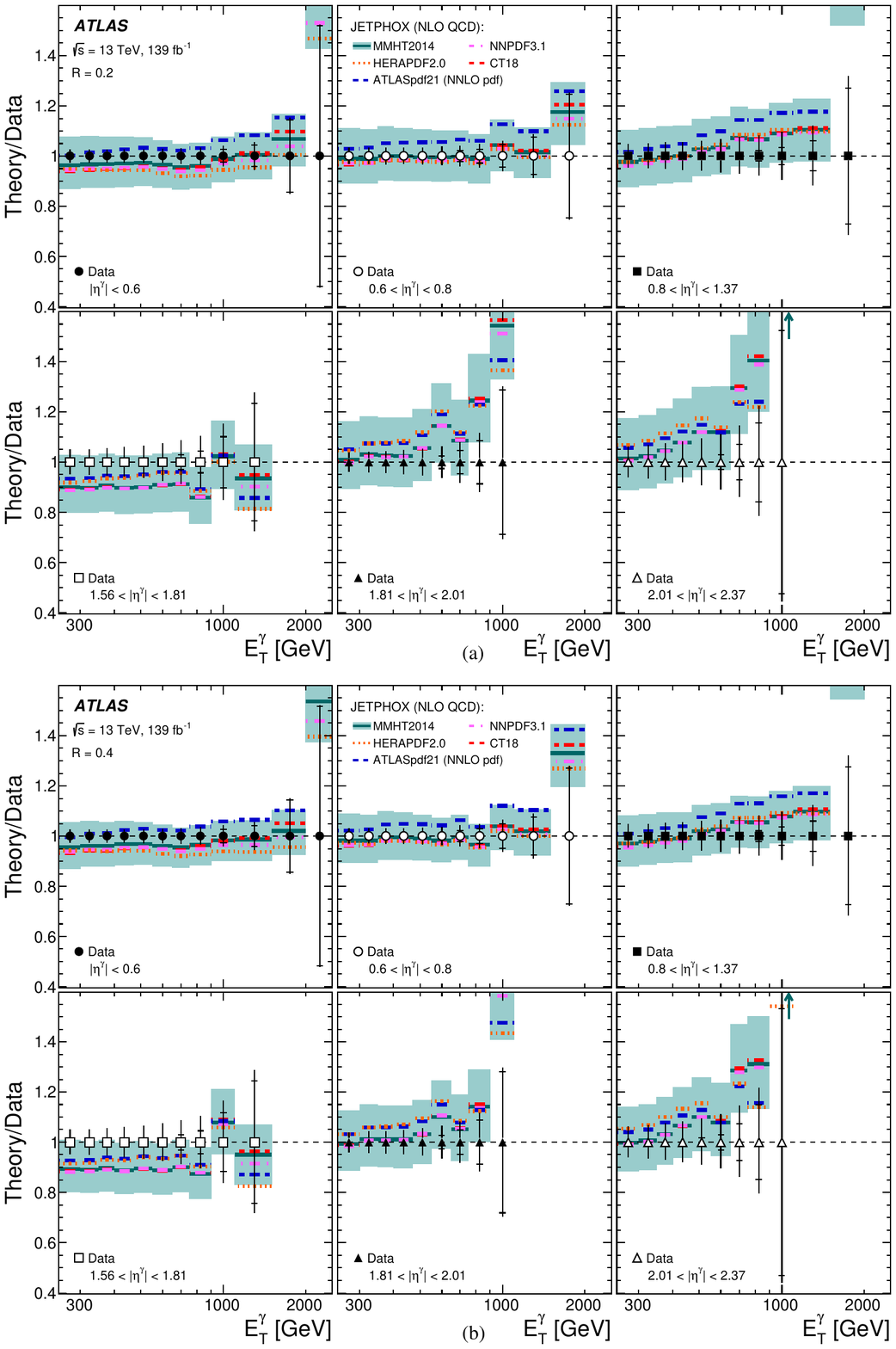}}
\end{picture}
\caption{
Ratio of the NLO pQCD calculations from \jetp\ based on
different PDF sets and the measured differential cross sections for
inclusive isolated-photon production with $R=0.2$ (a) and
$R=0.4$ (b) as functions of $\etg$ in different regions of $\etag$.
The inner (outer) error bars represent the statistical uncertainties
(statistical and systematic uncertainties added in quadrature). For
most of the points, the inner error bars are smaller than the marker size
and, thus, not visible. The shaded bands represent the theoretical uncertainty.
The arrows indicate the direction in which the ratios of
the calculations from \jetp\ and the measured differential cross
sections are located since they are outside of the plotted range in
these bins.
}
\label{fig13b}
\end{figure}

\begin{figure}[p]
\setlength{\unitlength}{1.0cm}
\begin{picture} (18.0,22.0)
\put (-3.9,-3.5){\includegraphics[width=23.1cm]{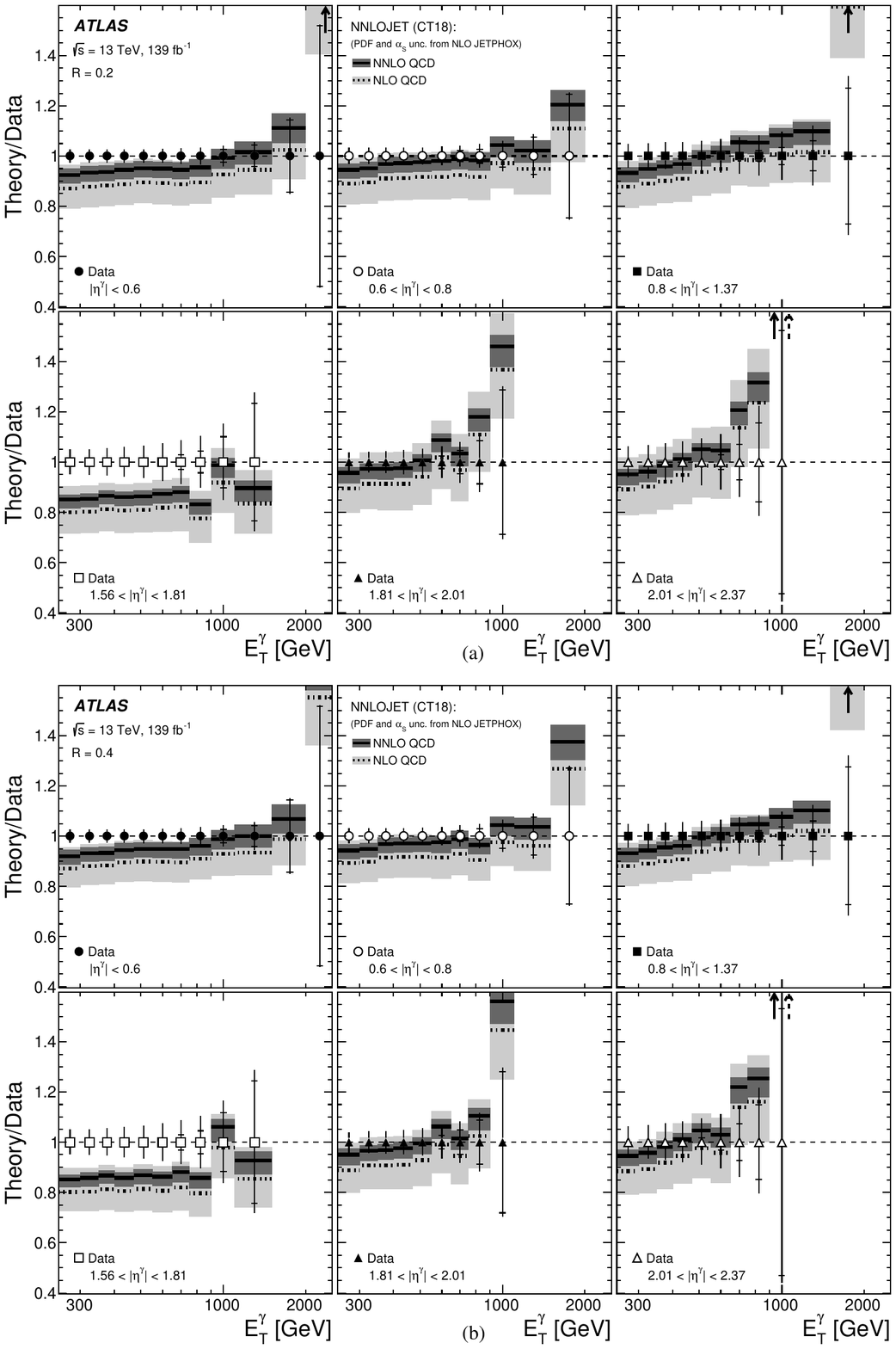}}
\end{picture}
\caption{
Ratio of the NLO (dotted lines) and NNLO (solid lines) pQCD
calculations from \nnlojet\ based on the CT18 PDF set and the measured
differential cross sections for isolated-photon production with $R=0.2$ (a) and
$R=0.4$ (b) as functions of $\etg$ in different regions of $\etag$.
The inner (outer) error bars represent the statistical uncertainties
(statistical and systematic uncertainties added in quadrature). For
most of the points, the inner error bars are smaller than the marker size
and, thus, not visible. The shaded bands represent the theoretical
uncertainties. The arrows indicate the direction in which the ratios of
the calculations from \nnlojet\ and the measured differential cross
sections are located since they are outside of the plotted range in
these bins.
}
\label{fig13_nnlojet}
\end{figure}

The ratio of the predictions from \sher\ NLO based on the NNPDF3.0 PDF
set and the measured cross sections is shown in Figure~\ref{fig13} and
the ratio of the predictions from \jetp\ based on different PDFs and
the measured cross sections is shown in Figure~\ref{fig13b}.
Both the predictions from \sher\ NLO and \jetp\ are consistent with the
measurements within the experimental and theoretical uncertainties.
However, the predictions of \sher\ NLO have a normalisation larger than
those of \jetp\ which is attributed to the fact that the former
include contributions from parton showers, virtual corrections for
$\gamma+{\textrm{2-jet}}$ and higher-order tree matrix elements for
the processes $2\rightarrow n$ with $n=4$ and $5$, which are not
present in the predictions of \jetp. As seen in Figure~\ref{fig13b}, the
\jetp\ predictions based on the MMHT2014, CT18 and NNPDF3.1 PDF sets are
similar and the closest to the data for $|\etag|<1.37$ and
$1.81<|\etag|<2.37$. For $1.56<|\etag|<1.81$, the predictions based on
the HERAPDF2.0 PDF and ATLASpdf21 sets are the closest to the data.
Figure~\ref{fig13_nnlojet} shows the ratio of the NLO and NNLO
predictions from \nnlojet\ based on the CT18 PDF set and the measured
cross sections; the predictions are consistent with the measurements
within the experimental and theoretical uncertainties, except in the region
$1.56<|\etag|<1.81$, where the NNLO pQCD predictions underestimate the
data.
 
\clearpage
 
\FloatBarrier
 
\newpage
 
\thispagestyle{empty}
 
\clearpage
 
\FloatBarrier
 
\newpage
 
\thispagestyle{empty}
 
\clearpage
 
\FloatBarrier
 
\newpage
 
\thispagestyle{empty}
 
\FloatBarrier
 
\clearpage
 
\newpage
 
\subsection{$R$ dependence of the fiducial cross
section for inclusive isolated-photon production}
\begin{figure}[h]
\setlength{\unitlength}{1.0cm}
\begin{picture} (18.0,12.5)
\put (-3.9,-14.0){\includegraphics[width=23.1cm]{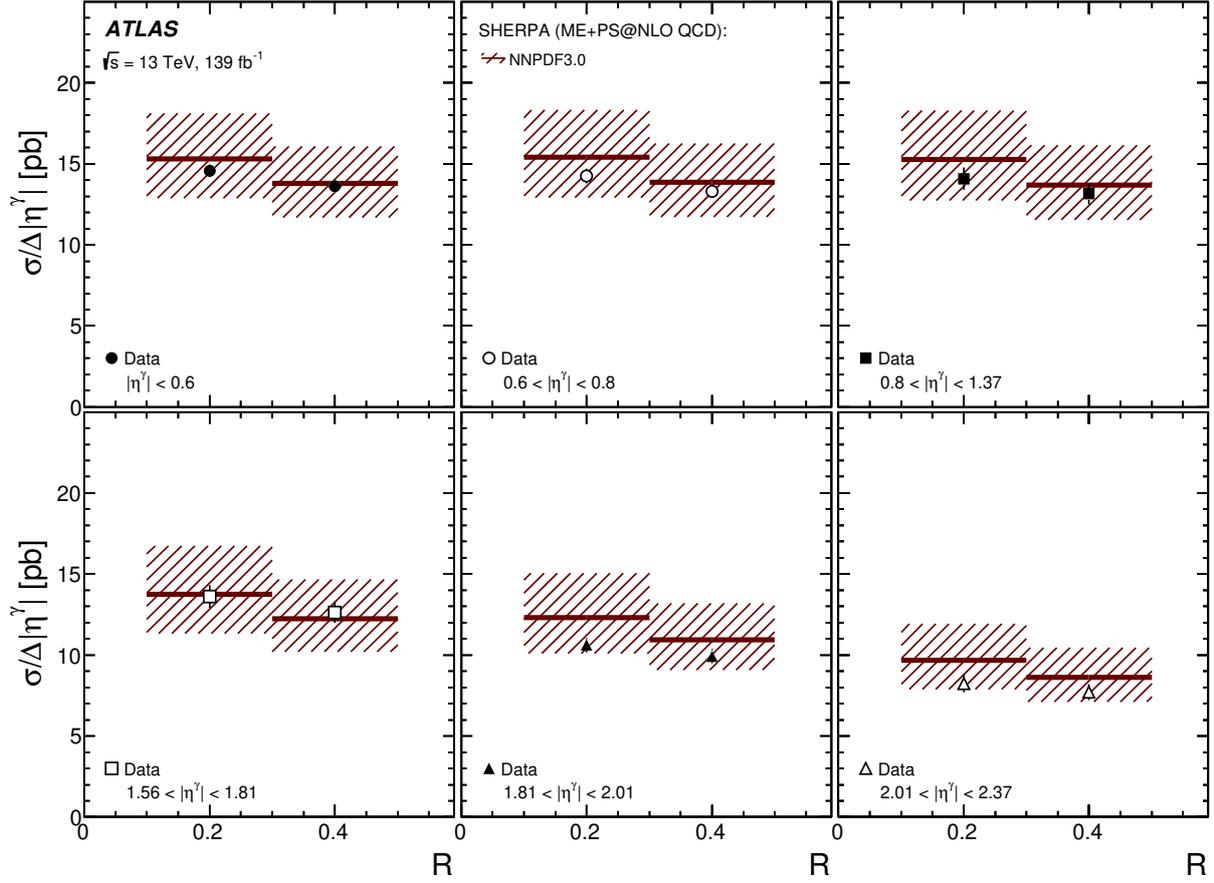}}
\end{picture}
\caption{
Measured fiducial integrated cross sections for
inclusive isolated-photon production as functions of $R$ in
different $\etag$ regions. The NLO pQCD predictions from \sher\ NLO based
on the NNPDF3.0 PDF set are also shown. The
error bars represent the statistical and systematic uncertainties added in
quadrature. For some of the points, the error bars are smaller than
the marker size and, thus, not visible. The hatched bands
represent the theoretical uncertainties.
}
\label{fig14}
\end{figure}

The dependence of the inclusive isolated-photon cross section
on $R$ is investigated by measuring the fiducial integrated cross
section in each $\etag$ region, divided by the width of the $|\etag|$
region, for both $R$ values measured (see Figure~\ref{fig14}).
The measured cross section decreases with increasing $R$ in all $\etag$
regions and it is approximately constant for $|\etag|<1.37$, but
decreases with increasing $\etag$ in the region $|\etag|>1.37$
for a fixed value of $R$.
 
\begin{figure}[h]
\setlength{\unitlength}{1.0cm}
\begin{picture} (18.0,12.5)
\put (-3.9,-14.0){\includegraphics[width=23.1cm]{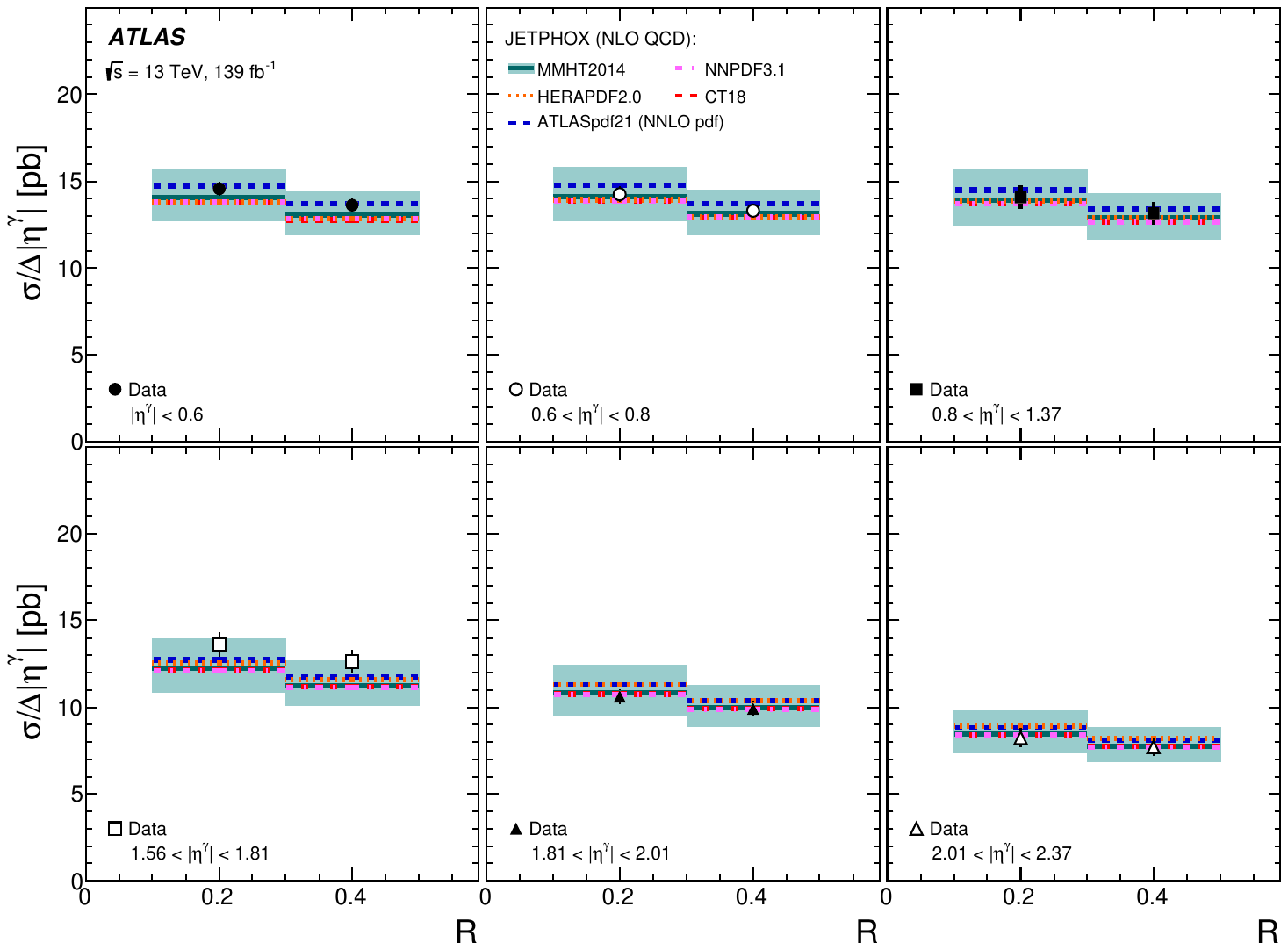}}
\end{picture}
\caption{
Measured fiducial integrated cross sections for
inclusive isolated-photon production as functions of $R$ in
different $\etag$ regions. The NLO pQCD predictions from
\jetp\ based on different PDF sets are also shown. The
error bars represent the statistical and systematic uncertainties added in
quadrature. For some of the points, the error bars are smaller than
the marker size and, thus, not visible. The shaded bands
represent the theoretical uncertainties.
}
\label{fig14b}
\end{figure}

The NLO pQCD predictions of \sher\ NLO and \jetp\ are compared to the data
in Figures~\ref{fig14} and \ref{fig14b}, respectively, and describe
within the theoretical and experimental uncertainties the dependence
on $R$ of the measured fiducial integrated cross sections. In
particular, the nominal predictions of \sher\ NLO tend to be above the
data for $R=0.2$ and $|\etag|<1.37$; in the region
$1.56<|\etag|<1.81$, these predictions describe the data well, but
there is a tendency to overestimate the data for $|\etag|>1.81$ for
both radii. The NLO pQCD predictions of \jetp\ describe the data well,
except in the region $1.56<|\etag|<1.81$, where the nominal prediction
of \jetp\ is below the data. Figure~\ref{fig14b} also includes the
\jetp\ predictions based on different PDFs; no significant sensitivity
to the PDFs is observed for the fiducial integrated cross sections.
 
\begin{figure}[p]
\setlength{\unitlength}{1.0cm}
\begin{picture} (18.0,12.5)
\put (-3.9,-14.0){\includegraphics[width=23.1cm]{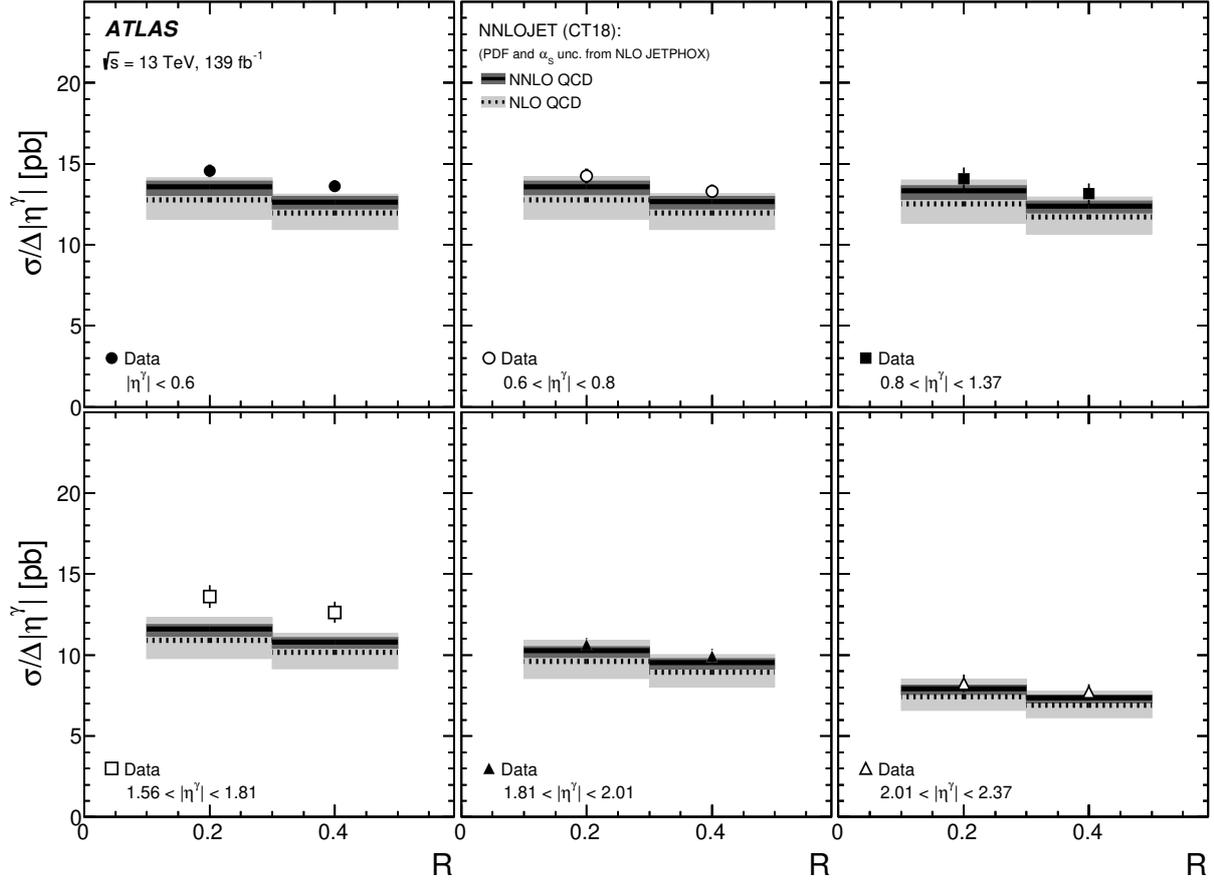}}
\end{picture}
\caption{
Measured fiducial integrated cross sections for isolated-photon production as
functions of $R$ in different $\etag$ regions. The NLO (dotted lines) and
NNLO (solid lines) pQCD predictions from \nnlojet\ based on the CT18 PDF set
are also shown. The
error bars represent the statistical and systematic uncertainties added in
quadrature. For some of the points, the error bars are smaller than
the marker size and, thus, not visible. The shaded bands represent the
theoretical uncertainties.
}
\label{fig14_nnlojet}
\end{figure}

Figure~\ref{fig14_nnlojet} shows the comparison of the measured
fiducial integrated cross section as a function of $R$ and the
predictions from \nnlojet. The NNLO pQCD predictions describe within the
theoretical and experimental uncertainties the dependence on $R$ of
the measured fiducial cross section, except in the region
$1.56<|\etag|<1.81$, where the predictions underestimate the data; for
$|\etag|<1.37$, there is a tendency in the NNLO pQCD predictions to be
below the data.
 
\FloatBarrier
 
\clearpage
 
\subsection{Ratio of the differential cross
sections with different isolation-cone radii}
Further investigation of the dependence on $R$ of the inclusive
isolated-photon cross sections is performed by measuring the ratios of
the differential cross sections for $R=0.2$ and $R=0.4$ as functions
of $\etg$ in different regions of $\etag$. For these measurements,
both the experimental, except for the $\etiso$ modelling (see
Section~\ref{etiso_mod}), and the theoretical uncertainties are
considered to be fully correlated. Thus, a significant cancellation of
the uncertainties is obtained in the ratio (see
Sections~\ref{totsys} and \ref{thunc}).
 
\begin{figure}[h]
\setlength{\unitlength}{1.0cm}
\begin{picture} (18.0,12.5)
\put (-3.9,-14.0){\includegraphics[width=23.1cm]{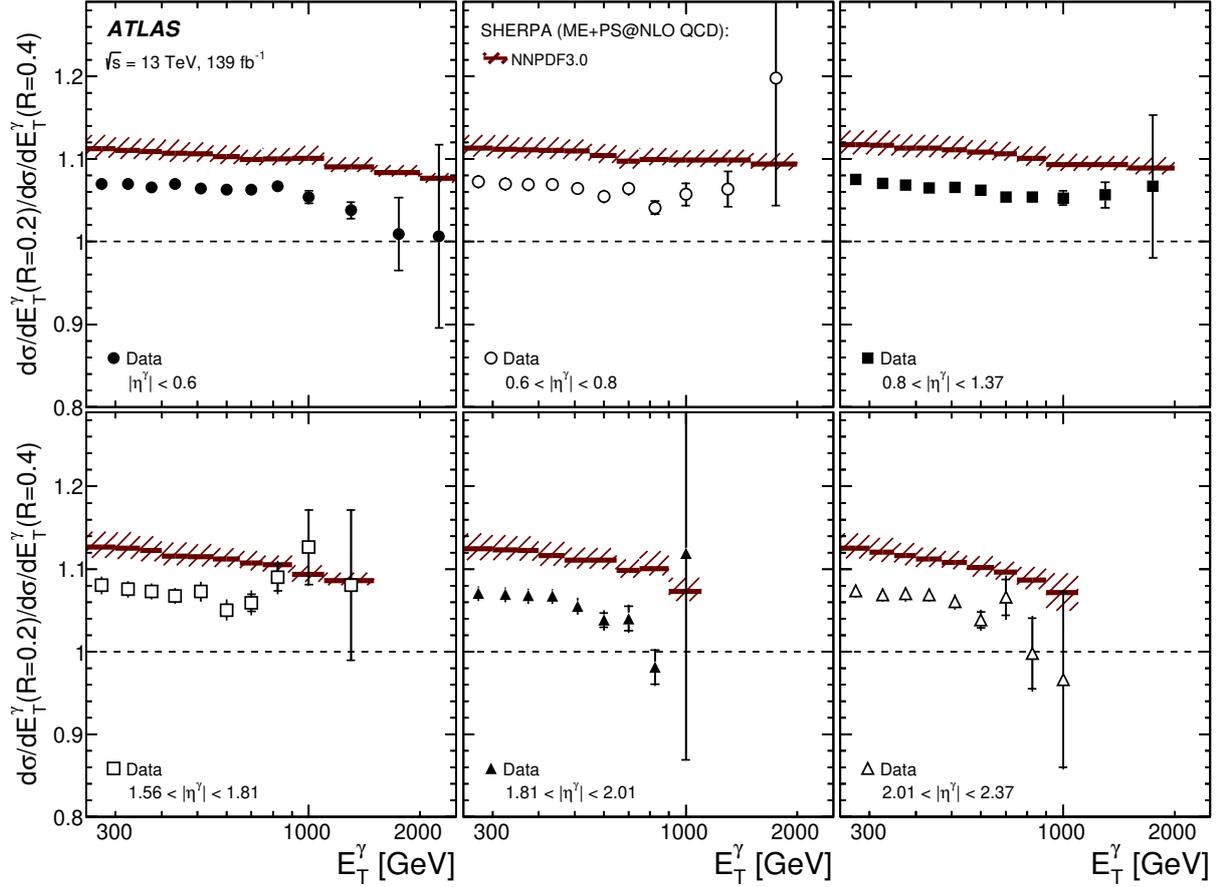}}
\end{picture}
\caption{
Measured ratios of the differential cross sections for inclusive
isolated-photon production for $R=0.2$ and $R=0.4$ as functions of
$\etg$ in different $\etag$ regions. The NLO pQCD predictions from
\sher\ NLO based on the NNPDF3.0 PDF set are also shown.
The inner (outer) error bars represent the statistical uncertainties
(statistical and systematic uncertainties added in quadrature) and
the hatched bands represent the theoretical uncertainty.
For some of the points, the inner and outer error bars are smaller than
the marker size and, thus, not visible.
}
\label{fig15}
\end{figure}

\begin{figure}[h]
\setlength{\unitlength}{1.0cm}
\begin{picture} (18.0,12.5)
\put (-3.9,-14.0){\includegraphics[width=23.1cm]{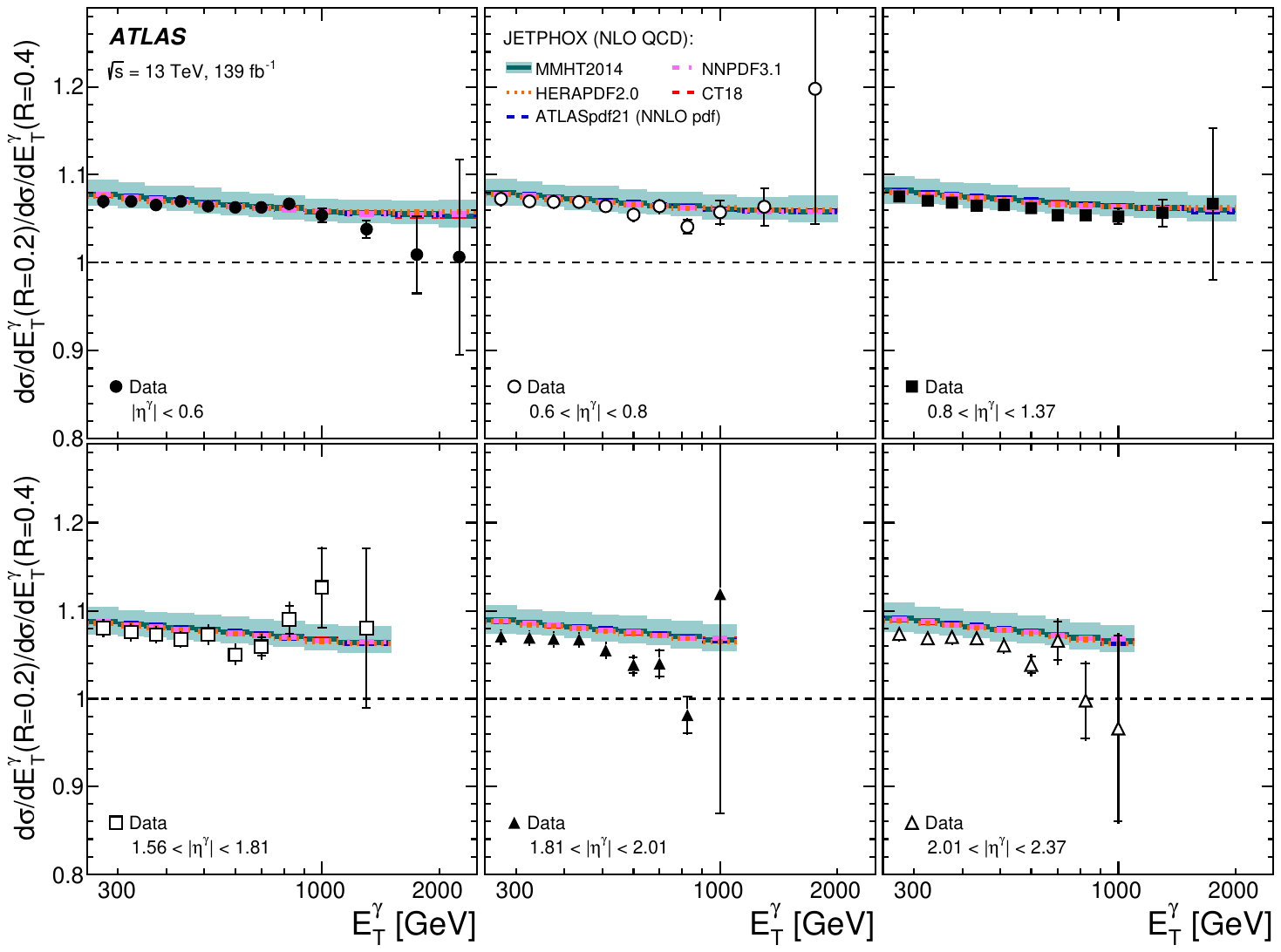}}
\end{picture}
\caption{
Measured ratios of the differential cross sections for inclusive
isolated-photon production for $R=0.2$ and $R=0.4$ as functions of
$\etg$ in different $\etag$ regions. The NLO pQCD predictions from
\jetp\ based on different PDF sets are also shown. The inner (outer)
error bars represent the statistical uncertainties (statistical and
systematic uncertainties added in quadrature) and the shaded bands
represent the theoretical uncertainties.
For some of the points, the inner and outer error bars are smaller than
the marker size and, thus, not visible.
}
\label{fig15b}
\end{figure}

\begin{figure}[p]
\setlength{\unitlength}{1.0cm}
\begin{picture} (18.0,12.5)
\put (-3.9,-14.0){\includegraphics[width=23.1cm]{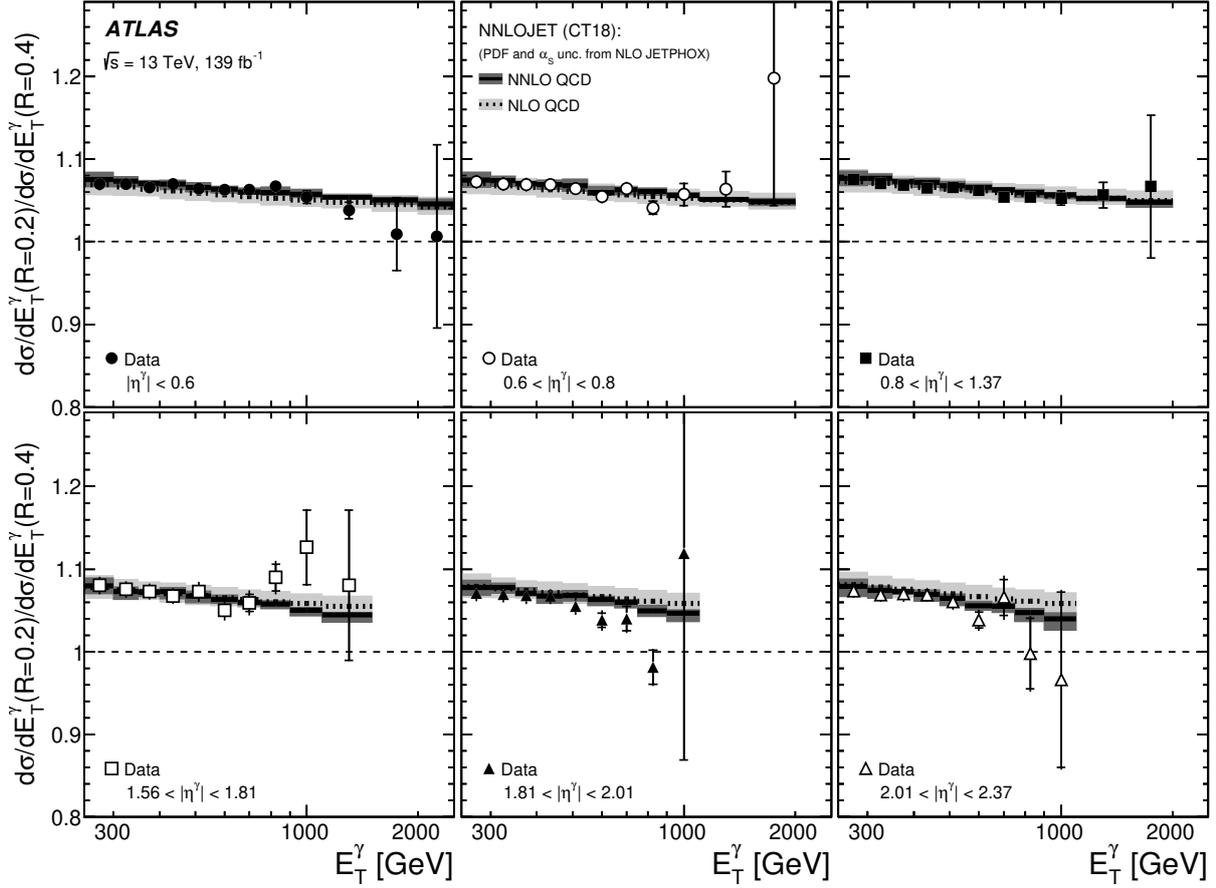}}
\end{picture}
\caption{
Measured ratios of the differential cross sections for inclusive
isolated-photon production for $R=0.2$ and $R=0.4$ as functions of
$\etg$ in different $\etag$ regions. The NLO (dotted lines) and
NNLO (solid lines) pQCD predictions from \nnlojet\ based on the CT18 PDF set
are also shown. The inner (outer) error bars represent the statistical
uncertainties (statistical and systematic uncertainties added in
quadrature) and the shaded bands represent the
theoretical uncertainties.
For some of the points, the inner and outer error bars are smaller than
the marker size and, thus, not visible.
}
\label{fig15_nnlojet}
\end{figure}

Figures~\ref{fig15} and \ref{fig15b} show the measured ratios together
with the predictions of \sher\ NLO and \jetp, respectively. The
measurements decrease with increasing $\etg$ in all $\etag$ regions
and have approximately the same value in all $\etag$ regions for a
fixed $\etg$ range. In the high-$\etg$ region, statistical
fluctuations distort this tendency in some $\etag$ regions. The NLO
pQCD predictions of \sher\ NLO overestimate the data in all $\etg$ and
$\etag$ regions, whereas those from \jetp\ give a good description of
the data. These differences between the predictions of \jetp\ and
\sher\ NLO might be attributed to the fact that the former includes an
explicit calculation of the fragmentation contribution using
fragmentation functions, an approach that describes the measurements
better.
No significant dependence on the proton PDFs is
observed in the ratios. Figure~\ref{fig15_nnlojet} shows the measured
ratios together with the predictions of \nnlojet. The NNLO pQCD
predictions give a good description of the data. These measurements
provide a very stringent test of pQCD with reduced experimental and
theoretical uncertainties.
 
\FloatBarrier
 
\section{Summary and conclusions}
\label{conc}
Measurements of the inclusive isolated-photon production cross
sections in $pp$ collisions at $\sqrt s=13$~\TeV\ recorded by the
ATLAS experiment at the LHC are presented based on $139$~fb$^{-1}$ of
2015--2018 data.
 
Differential cross sections as functions of $\etg$ are measured
in different regions of $\etag$ for photons with $\etg>250$~\GeV\ and
$|\etag|<2.37$, excluding the region $1.37<|\etag|<1.56$.
The photon isolation is ensured by requiring that the transverse
energy in a cone of $R=0.2$ or $R=0.4$ around the photon direction
is smaller than \etisocut. Values of $\etg$ up to $2.5$~\TeV\ are
measured with the full Run 2 ATLAS data set.
 
The measurements presented in this paper constitute an improvement
with respect to those published earlier in several aspects. The
$\etag$ range is subdivided in more regions; this provides more
detailed experimental information for the PDF fits. The measurements
are performed based on different isolation-cone radii, namely $R=0.2$
and $R=0.4$, which provide a test of the dependence of the pQCD
predictions on $R$; these tests are performed in terms of the fiducial
integrated cross section as functions of $R$ in different regions of
$\etag$ and of the ratio of the differential cross sections for
$R=0.2$ and $R=0.4$ as functions of $\etg$ in different regions of
$\etag$.
 
Next-to-leading-order pQCD predictions using several PDF sets are
compared to the differential cross section measurements and found to
provide an adequate description of the data within the experimental
and theoretical uncertainties. The comparison of data and theory is
limited by the theoretical uncertainties due to missing higher-order
terms in pQCD; in particular, the predictions from \sher\ NLO have a tendency to be
above the data whereas the predictions from \jetp\ provide a good
description of the data in all $\etag$ for both isolation-cone radii.
Experimental systematic uncertainties are smaller than the theoretical
uncertainties over the full investigated phase space. The measurements
have the potential to further constrain the PDFs, particularly the
gluon density in the proton, within a global NNLO QCD fit.
 
The dependence on $R$ of the measured cross section for inclusive
isolated-photon production is described well by the predictions of
\jetp, whereas the predictions of \sher\ NLO for the ratios are above the
data in most of the $\etag$ and $\etg$ regions. No dependence
on the proton PDFs of the predictions for the fiducial cross section as
functions of $R$ or the ratio of the differential cross sections with $R=0.2$
and $R=0.4$ is observed. These ratios provide a very stringent test of
pQCD, with significantly reduced experimental and theoretical
uncertainties, and validate the underlying theoretical description up
to $\oas$.
 
Next-to-next-to-leading-order pQCD predictions, including direct and
fragmentation components, are compared to the differential and
fiducial cross sections and to the ratios of the cross sections. For
both cone radii, the NNLO predictions give a good description of the
data within the uncertainties, except in the region
$1.56<|\etag|<1.81$, where the calculations underestimate the
data. The comparison of the ratios of the differential cross sections
between data and the predictions including NNLO corrections
validates the underlying pQCD theoretical description up to $\oass$.
 
\section*{Acknowledgements}


We thank CERN for the very successful operation of the LHC, as well as the
support staff from our institutions without whom ATLAS could not be
operated efficiently.
 
We acknowledge the support of
ANPCyT, Argentina;
YerPhI, Armenia;
ARC, Australia;
BMWFW and FWF, Austria;
ANAS, Azerbaijan;
CNPq and FAPESP, Brazil;
NSERC, NRC and CFI, Canada;
CERN;
ANID, Chile;
CAS, MOST and NSFC, China;
Minciencias, Colombia;
MEYS CR, Czech Republic;
DNRF and DNSRC, Denmark;
IN2P3-CNRS and CEA-DRF/IRFU, France;
SRNSFG, Georgia;
BMBF, HGF and MPG, Germany;
GSRI, Greece;
RGC and Hong Kong SAR, China;
ISF and Benoziyo Center, Israel;
INFN, Italy;
MEXT and JSPS, Japan;
CNRST, Morocco;
NWO, Netherlands;
RCN, Norway;
MEiN, Poland;
FCT, Portugal;
MNE/IFA, Romania;
MESTD, Serbia;
MSSR, Slovakia;
ARRS and MIZ\v{S}, Slovenia;
DSI/NRF, South Africa;
MICINN, Spain;
SRC and Wallenberg Foundation, Sweden;
SERI, SNSF and Cantons of Bern and Geneva, Switzerland;
MOST, Taiwan;
TENMAK, T\"urkiye;
STFC, United Kingdom;
DOE and NSF, United States of America.
In addition, individual groups and members have received support from
BCKDF, CANARIE, Compute Canada and CRC, Canada;
PRIMUS 21/SCI/017 and UNCE SCI/013, Czech Republic;
COST, ERC, ERDF, Horizon 2020 and Marie Sk{\l}odowska-Curie Actions, European Union;
Investissements d'Avenir Labex, Investissements d'Avenir Idex and ANR, France;
DFG and AvH Foundation, Germany;
Herakleitos, Thales and Aristeia programmes co-financed by EU-ESF and the Greek NSRF, Greece;
BSF-NSF and MINERVA, Israel;
Norwegian Financial Mechanism 2014-2021, Norway;
NCN and NAWA, Poland;
La Caixa Banking Foundation, CERCA Programme Generalitat de Catalunya and PROMETEO and GenT Programmes Generalitat Valenciana, Spain;
G\"{o}ran Gustafssons Stiftelse, Sweden;
The Royal Society and Leverhulme Trust, United Kingdom.
 
The crucial computing support from all WLCG partners is acknowledged gratefully, in particular from CERN, the ATLAS Tier-1 facilities at TRIUMF (Canada), NDGF (Denmark, Norway, Sweden), CC-IN2P3 (France), KIT/GridKA (Germany), INFN-CNAF (Italy), NL-T1 (Netherlands), PIC (Spain), ASGC (Taiwan), RAL (UK) and BNL (USA), the Tier-2 facilities worldwide and large non-WLCG resource providers. Major contributors of computing resources are listed in Ref.~\cite{ATL-SOFT-PUB-2021-003}.


\printbibliography
 
\clearpage
\input{atlas_authlist}

\clearpage

\end{document}

%% file: atlas_authlist.tex
 
\begin{flushleft}
\hypersetup{urlcolor=black}
{\Large The ATLAS Collaboration}

\bigskip

\AtlasOrcid[0000-0002-6665-4934]{G.~Aad}$^\textrm{\scriptsize 102}$,
\AtlasOrcid[0000-0002-5888-2734]{B.~Abbott}$^\textrm{\scriptsize 120}$,
\AtlasOrcid[0000-0002-1002-1652]{K.~Abeling}$^\textrm{\scriptsize 55}$,
\AtlasOrcid[0000-0002-8496-9294]{S.H.~Abidi}$^\textrm{\scriptsize 29}$,
\AtlasOrcid[0000-0002-9987-2292]{A.~Aboulhorma}$^\textrm{\scriptsize 35e}$,
\AtlasOrcid[0000-0001-5329-6640]{H.~Abramowicz}$^\textrm{\scriptsize 151}$,
\AtlasOrcid[0000-0002-1599-2896]{H.~Abreu}$^\textrm{\scriptsize 150}$,
\AtlasOrcid[0000-0003-0403-3697]{Y.~Abulaiti}$^\textrm{\scriptsize 117}$,
\AtlasOrcid[0000-0003-0762-7204]{A.C.~Abusleme~Hoffman}$^\textrm{\scriptsize 137a}$,
\AtlasOrcid[0000-0002-8588-9157]{B.S.~Acharya}$^\textrm{\scriptsize 69a,69b,p}$
% ...
\bigskip
\\

% ...
$^{29}$Physics Department, Brookhaven National Laboratory, Upton NY; United States of America.\\
% ...
$^{35}$$^{(a)}$Facult\'e des Sciences Ain Chock, R\'eseau Universitaire de Physique des Hautes Energies - Universit\'e Hassan II, Casablanca;$^{(b)}$Facult\'{e} des Sciences, Universit\'{e} Ibn-Tofail, K\'{e}nitra;$^{(c)}$Facult\'e des Sciences Semlalia, Universit\'e Cadi Ayyad, LPHEA-Marrakech;$^{(d)}$LPMR, Facult\'e des Sciences, Universit\'e Mohamed Premier, Oujda;$^{(e)}$Facult\'e des sciences, Universit\'e Mohammed V, Rabat;$^{(f)}$Institute of Applied Physics, Mohammed VI Polytechnic University, Ben Guerir; Morocco.\\
% ...
$^{55}$II. Physikalisches Institut, Georg-August-Universit\"{a}t G\"ottingen, G\"ottingen; Germany.\\
% ...
$^{69}$$^{(a)}$INFN Gruppo Collegato di Udine, Sezione di Trieste, Udine;$^{(b)}$ICTP, Trieste;$^{(c)}$Dipartimento Politecnico di Ingegneria e Architettura, Universit\`a di Udine, Udine; Italy.\\
% ...
$^{102}$CPPM, Aix-Marseille Universit\'e, CNRS/IN2P3, Marseille; France.\\
% ...
$^{117}$Department of Physics, New York University, New York NY; United States of America.\\
% ...
$^{120}$Homer L. Dodge Department of Physics and Astronomy, University of Oklahoma, Norman OK; United States of America.\\
% ...
$^{137}$$^{(a)}$Departamento de F\'isica, Pontificia Universidad Cat\'olica de Chile, Santiago;$^{(b)}$Millennium Institute for Subatomic physics at high energy frontier (SAPHIR), Santiago;$^{(c)}$Instituto de Investigaci\'on Multidisciplinario en Ciencia y Tecnolog\'ia, y Departamento de F\'isica, Universidad de La Serena;$^{(d)}$Universidad Andres Bello, Department of Physics, Santiago;$^{(e)}$Instituto de Alta Investigaci\'on, Universidad de Tarapac\'a, Arica;$^{(f)}$Departamento de F\'isica, Universidad T\'ecnica Federico Santa Mar\'ia, Valpara\'iso; Chile.\\
% ...
$^{150}$Department of Physics, Technion, Israel Institute of Technology, Haifa; Israel.\\
$^{151}$Raymond and Beverly Sackler School of Physics and Astronomy, Tel Aviv University, Tel Aviv; Israel.\\
% ...
$^{a}$ Also Affiliated with an institute covered by a cooperation agreement with CERN.\\
$^{b}$ Also at An-Najah National University, Nablus; Palestine.\\
$^{c}$ Also at Borough of Manhattan Community College, City University of New York, New York NY; United States of America.\\
$^{d}$ Also at Bruno Kessler Foundation, Trento; Italy.\\
$^{e}$ Also at Center for High Energy Physics, Peking University; China.\\
$^{f}$ Also at Center for Interdisciplinary Research and Innovation (CIRI-AUTH), Thessaloniki ; Greece.\\
% ...
$^{p}$ Also at Department of Physics, King's College London, London; United Kingdom.\\
% ...

\end{flushleft}
